\documentclass[traditabstract]{aa} 
\bibpunct{(}{)}{;}{a}{}{,}

\newcommand{\ergps}{erg\thinspace s$^{-1}$}

\newcommand{\ergpspsqcm}{erg\thinspace s$^{-1}$\thinspace cm$^{-2}$}
\newcommand{\psqcm}{cm$^{-2}$}
\newcommand{\nH}{$N_{\rm H}$}

\newcommand{\Msunpyr}{$M_{\odot}$\thinspace yr$^{-1}$}

\usepackage[varg]{txfonts}
\usepackage{graphicx}
\usepackage{subfig}
\begin{document}
\title{X-ray supernova remnants in the starburst region of M82}


\author{K. Iwasawa\inst{1,2}
}

\institute{Institut de Ci\`encies del Cosmos (ICCUB), Universitat de Barcelona (IEEC-UB), Mart\'i i Franqu\`es, 1, 08028 Barcelona, Spain
         \and
ICREA, Pg. Llu\'is Companys 23, 08010 Barcelona, Spain
}


 
\abstract{
We searched for X-ray supernova remnants (SNRs) in the starburst region of M82, using archival data from the {\it Chandra X-ray Observatory} with a total effective exposure time of 620 ks with an X-ray spectroscopic selection. Strong line-emission from Fe {\sc xxv} at 6.7 keV is a characteristic spectral feature of hot, shocked gas of young SNRs and distinctive among the discrete sources in the region populated by X-ray binaries. We selected candidates using narrow-band imaging aimed at the line excess and identified six (and possibly a seventh) X-ray SNRs. Two previously known examples were recovered by our selection. Five of them have radio counterparts, including the radio supernova SN2008iz, which was discovered as a radio transient in 2008. It shows a hard X-ray spectrum with a blueshifted Fe K feature with $v\approx -2700$ km\thinspace s$^{-1}$, both of which suggest its youth. The 4-8 keV luminosities of the selected SNRs are in the range of (0.3-2.5)$\times 10^{38}$ \ergps. We made a crude estimate of the supernova rate, assuming that more luminous SNRs are younger, and found $\nu_{\rm SN}=0.06$ (0.03-0.13) yr$^{-1}$, in agreement with the supernova rates estimated by radio observations and the generally believed star formation rate of M82, although the validity of the assumption is questionable. A sum of the Fe~{\sc xxv} luminosity originating from the selected X-ray SNRs consists of half of the total Fe~{\sc xxv} luminosity observed in the central region of M82. We briefly discuss its implications for starburst winds and the Fe~{\sc xxv} emission in more luminous starburst galaxies.}

\keywords{X-rays: galaxies - Galaxies: starburst - Galaxies:individual (M82) - ISM: supernova remnants}
\titlerunning{X-ray SNRs in M82}
\authorrunning{K. Iwasawa}
\maketitle
%

\section{Introduction}

M82 is a well-studied, nearby starburst galaxy with a star formation rate (SFR) of $\sim 10$\Msunpyr\ \citep[e.g.][]{degrijs01}. At such a high SFR, core-collapse supernovae (hereafter we call them simply supernovae or SNe, unless stated otherwise) are expected to occur frequently and supernova remnants (SNRs) are accumulated over a starburst duration. However, they reside in a region of heavy dust obscuration which hinders observations at optical wavelengths. High-resolution radio imaging has instead revealed a population of radio SNRs in the central star forming disc extending over $45^{\prime\prime}$ ($\sim 700$ pc) in M82 \citep[e.g.][]{kronberg85}.

Analogously, young SNRs can be detected at X-ray energies above a few keV, where the severe effect of obscuration is reduced. The {\it Chandra X-ray Observatory (Chandra)} provides a sub-arcsecond resolution \citep{weisskopf02} which would be sufficient for resolving young SNRs in the starburst disc in M82. There are a number of {\it Chandra} observations of M82 in the archive and $\sim 600$ ks of exposure time for the central region of M82 with the ACIS camera is available. Hot, diffuse X-ray emission has been found to envelope the starburst region \citep{griffiths00,strickland07,liu14} where SNRs are likely embedded. This diffuse emission induces high background noise, limiting the source detection effciency, but the youngest SNRs emitting with an X-ray luminosity of $\sim 10^{38}$ \ergps\ would be detectable if the full exposure of the {\it Chandra} datasets were used. With a SFR of 10\Msunpyr, one SN is expected to occur every $\sim 15$ yr \citep{heckman90}. This means, for instance, about six SNRs that are younger than 100 yr old are expected to be present.  

Among known SNRs in the Milky Way, apart from some synchrotron-dominated SNRs such as G1.9+03 \citep{reynolds08}, X-ray emission from young SNRs is dominated by thermal emission from a shocked medium and its spectrum exhibits a strong Fe K$\alpha $ emission, as seen in Cas A \citep{holt94,kaastra95,vink96}. Although Cas A is about 350 yr old and its Fe K line is now found at $\simeq 6.5$ keV, as the medium shocked by a blast wave has cooled, younger SNRs are expected to strongly emit Fe {\sc xxv} emission at 6.7 keV, as seen in the X-ray spectrum of SN1993J \citep{kohmura94,uno02}. Although abundant discrete X-ray sources in the starburst region of M82 are dominated by X-ray binaries, this spectral feature is unique to young SNRs and is a powerful tool for distinguishing them from X-ray binaries. There has been one {\it Chandra}-detected X-ray source with Fe {\sc xxv} in the central region of M82, as reported by \citet[][their Source \#18]{gandhi11} and \citet[][their Source A]{liu14}. It is associated with the radio source 44.01+59.6, classified as a SNR \citep{mcdonald02}. Another X-ray source with a thermal emission spectrum has also been identified with the radio SNR 41.95+57.5 \citep{kong07,chiang11,strickland07}, although Fe {\sc xxv} was not detected, due to the shorter exposure.  

In this work, we present a systematic search for obscured X-ray SNRs in the starburst region of M82 using archival {\it Chandra} data as selected by their Fe~{\sc xxv} emission and discuss their implications. This is an independent X-ray selection of SNRs, allied with the work in which SNRs in nearby galaxies were identified by soft X-ray spectra of the low-temperature component of evolved SNR emission \citep{leonidaki10,long10}. Our selection instead aims at the high-temperature component of young SNRs in the obscured starburst region where soft X-ray emission is suppressed. The distance to M82 is assumed to be 3.6 Mpc \citep{freedman94}. The angular scale is thus 17.4 pc arcsec$^{-1}$ (1 kpc corresponds to approximately 1 arcmin). 

\section{Observations}


\begin{table}
\caption{{\it Chandra} observations of M82.}
\label{tab:obslog}
\centering
\begin{tabular}{ccccccc}
  \hline\hline
Period & Date & ObsID & ACIS & Exp. & $\theta$ & $\phi$ \\
  \hline
  I & 1999-09-20 & 361 & I3 & 33 & 0.3 & 26 \\
  & 1999-09-20 & 1302 & I3 & 15 & 0.3 & 26\\
  & 2002-06-18 & 2933 & S3 & 18 & 0.6 & 1 \\[5pt]
  II & 2009-06-24 & 10542 & S3 & 118 & 1.4 & 223 \\
  & 2009-07-01 & 10543 & S3 & 118 & 1.1 & 179 \\
  & 2009-07-07 & 10925 & S3 & 45 & 2.6 & 21 \\
  & 2009-07-08 & 10544 & S3 & 74 & 2.6 & 21 \\[5pt]
  III & 2010-06-17 & 11104 & S3 & 10 & 0.4 & 50 \\
  & 2010-07-20 & 11800 & S3 & 17 & 2.7 & 1 \\
  & 2010-07-23 & 10545 & S3 & 95 & 2.7 & 8 \\[5pt]
  IV & 2012-08-09 & 13796 & S3 & 20 & 0.3 & 169 \\
  & 2014-02-03 & 16580 & S3 & 47 & 1.2 & 323 \\
  & 2015-01-20 & 16023 & S3 & 10 & 1.2 & 356 \\
\hline
\end{tabular}
\tablefoot{Period: the four separate periods of observations defined in the text. Date: the the date when each observation started. ACIS: the ACIS CCD chip on which the central part of the M82 was placed. Exp.: useful exposure time in units of $10^3$ seconds. $\theta $, $\phi $: Off-axis angle in arcminutes and roll angle in degrees of the central position of Fig. 1 in each observation.}
\end{table}

We used 13 {\it Chandra} ACIS observations between 1999 and 2015 listed in Table \ref{tab:obslog}. Three or four observations are clustered in each of the four periods, labelled as I (1999-2002), II (2009), III (2010), and IV (2012-2015). The central part of M82 was imaged with the ACIS-I3 in the earliest two observations and with the ACIS-S3 in the rest of the observations. The aim point of the telescope varies between the observations (see the off-axis angle $\theta $ of each observation in Table \ref{tab:obslog}), depending on the principal aim of each observation. This means the point spread functions (PSFs) of the region of our interest in some observations are more elongated than in others. A good PSF was maintained across the region of our interest during the two observations (ObsID 10542 and 10543) with the longest exposure. The event file of ObsID 10543 was used as the reference and all the other event files were re-projected onto it for the imaging analysis presented below. We mainly used data in the 4-8 keV band which includes the Fe K emission at 6.7 keV and is little affected by obscuration of a column density up to \nH $\sim 1\times 10^{23}$ \psqcm.

{\it Chandra} data reduction and image analysis were carried out by CIAO 4.12 \citep{fruscione06} using the calibration files in CALDB4.53. Spectral analysis was performed using HEASoft/XSPEC \citep{blackburn99,arnaud96}.

\section{Results}

\subsection{X-ray imaging of the central starburst region}


\begin{figure*}
\centerline{\includegraphics[width=0.95\textwidth,angle=0]{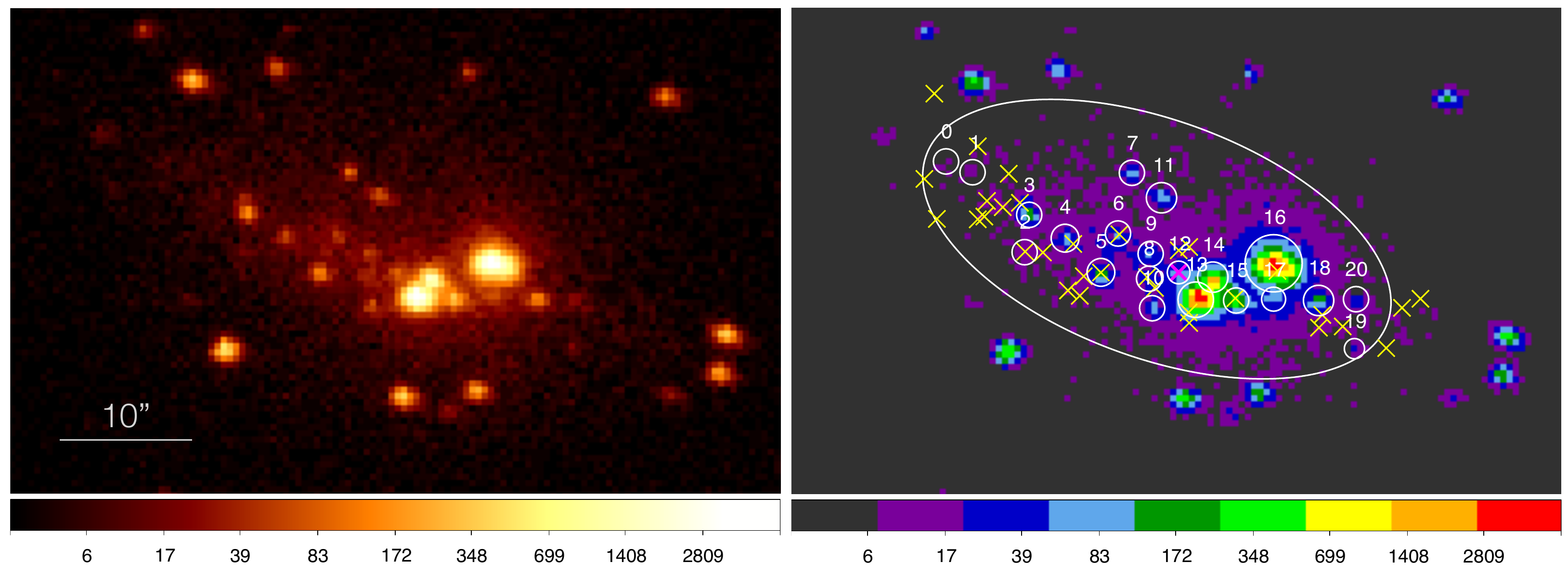}}
\caption{Left: {\it Chandra} 4-8 keV image of the central part of M82. All the 13 exposures are used to construct the image. The image is centred at RA$= 09^{\rm h}55^{\rm m}51.6^{\rm s}$, Dec. $+69^{\circ}40^{\prime}46^{\prime\prime}$. The image size is $60^{\prime\prime}\times 40^{\prime\prime}$. The scale bar indicates 10 arcsec (projected distance of $\approx 170$ pc at M82). Regarding the image orientation, north is up and east is to the left. Data are integrated counts displayed in logarithmic intervals. Right: Same image as in the left panel, but in nine discrete colour intervals (aips0) overlaid with an ellipse enclosing the bright part of the diffuse emission and the spectral aperture for 21 discrete sources detected within the ellipse. The source numbers from 0 to 20 assigned to individual sources are labelled. Positions of the 32 radio SNR candidates selected by \citet{fenech10} and of SN2008iz are marked by cross symbols in yellow and magenta, respectively.}
\label{fig:e48img}
\end{figure*}

Figure \ref{fig:e48img} shows the 4-8 keV band image of the central part ($60^{\prime\prime}\times 40^{\prime\prime}$) of M82, integrating all the 13 exposures of 620 ks. A number of discrete sources are embedded within the diffuse emission. This region is the base of the galactic-scale bi-conical winds imaged in various wavelengths \citep[e.g.][]{armus90,shopbell98,yoshida11,roussel10,watson84,strickland97}. The diffuse emission is elongated along the galactic plane with PA$\sim 70^{\circ}$. The ellipse in the image marks the brighter part of the diffuse emission.
Its major axis is $\approx 39^{\prime\prime}$ long and the enclosed region coincides with the area where SNRs were mapped at radio wavelengths. The discrete X-ray sources within this ellipse are tightly distributed along the galactic plane and many of them are likely to lie in the molecular disc where a starburst is taking place. Therefore they are likely to contain SNRs. As a result of a longer exposure time, more discrete sources appear to be seen, compared to the previously published {\it Chandra} images at similar energies \citep{strickland07,liu14}. We performed a source detection and selected sources within the ellipse since those located outside are likely X-ray binaries.
The limiting 4-8 keV luminosity for selected X-ray SNR candidates is estimated to be $0.9\times 10^{37}$ \ergps, which can vary slightly, depending on brightness of the surrounding diffuse background emission. In this estimate, we assumed a thermal emission spectrum of the Astrophysical Plasma Emission Code {\tt apec} \citep{foster12} with a temperature of $kT = 5.5$ keV.

\subsection{Discrete sources within the hot diffuse emission region}

We used {\tt celldetect} of CIAO for source detection. It works well for detecting point-like sources down to a low flux level despite the region being relatively crowded. The PSF map was created using the event file of ObsID~10543 in which the PSF is the sharpest among the observations of longer exposures. The use of this PSF map is justified by the fact that a source detection would be driven by the sharp core of the integrated image which is a mixture of the sharp and slightly elongated PSFs from various observations (see Sect. 2). We detected 21 sources within the hot diffuse ellipse with the detection threshold of $3\sigma $.
The 21 sources are indicated in Fig. \ref{fig:e48img} and the source number 0-20 were assigned in the descending order of RA (in the direction of east to west) of their positions in J2000. We call these X-ray point-like sources as Xp0 through Xp20. Corresponding official {\it Chandra} source names -- the majority of which have been registered in the {\it Chandra} Source Catalog 2.0 \citep{evans20} -- are listed in Appendix A. The 90\% uncertainty of {\it Chandra} positions has been estimated to be $0\farcs 8$ \footnote{https://cxc.harvard.edu/cal/ASPECT/celmon/}.

Many of the 21 sources are expected to be X-ray binaries. They include well-studied ultra-luminous X-ray sources (ULXs) M82 X-1 and X-2 \citep[e.g.][references therein]{kaaret01,brightman19} and transient flaring sources, which are clearly accreting compact objects. We pick SNR candidates below, however, solely by an X-ray spectroscopic selection without prior knowledge of source types. The rationale for the selection is as follows. A thermal emission spectrum of a temperature of a few keV expected from a young SNR exhibits a strong Fe~{\sc xxv} line with an equivalent width (EW) $\ga 0.5$ keV. This high-ionization Fe line is absent or very weak in spectra of X-ray binaries. Fe K emission in X-ray binaries, when it is present, is primarily a 6.4 keV cold line that is weaker and sometimes broadened \citep[e.g.][]{ueda09,revnivtsev99}. A 6.7 keV line could be visible only when the continuum is strongly suppressed, for example, in eclipse \citep{nagase94,miller20} but this state would be too faint to be detected in M82, and furthermore, a stronger 6.4 keV line is always present. Therefore, searching for sources showing excess in the Fe~{\sc xxv} band around 6.7 keV would select young SNR candidates and filter out X-ray binaries. 

\subsection{Narrow-band imaging selection of Fe {\sc xxv} sources}

\begin{figure}
  \centerline{\includegraphics[width=0.4\textwidth,angle=0]{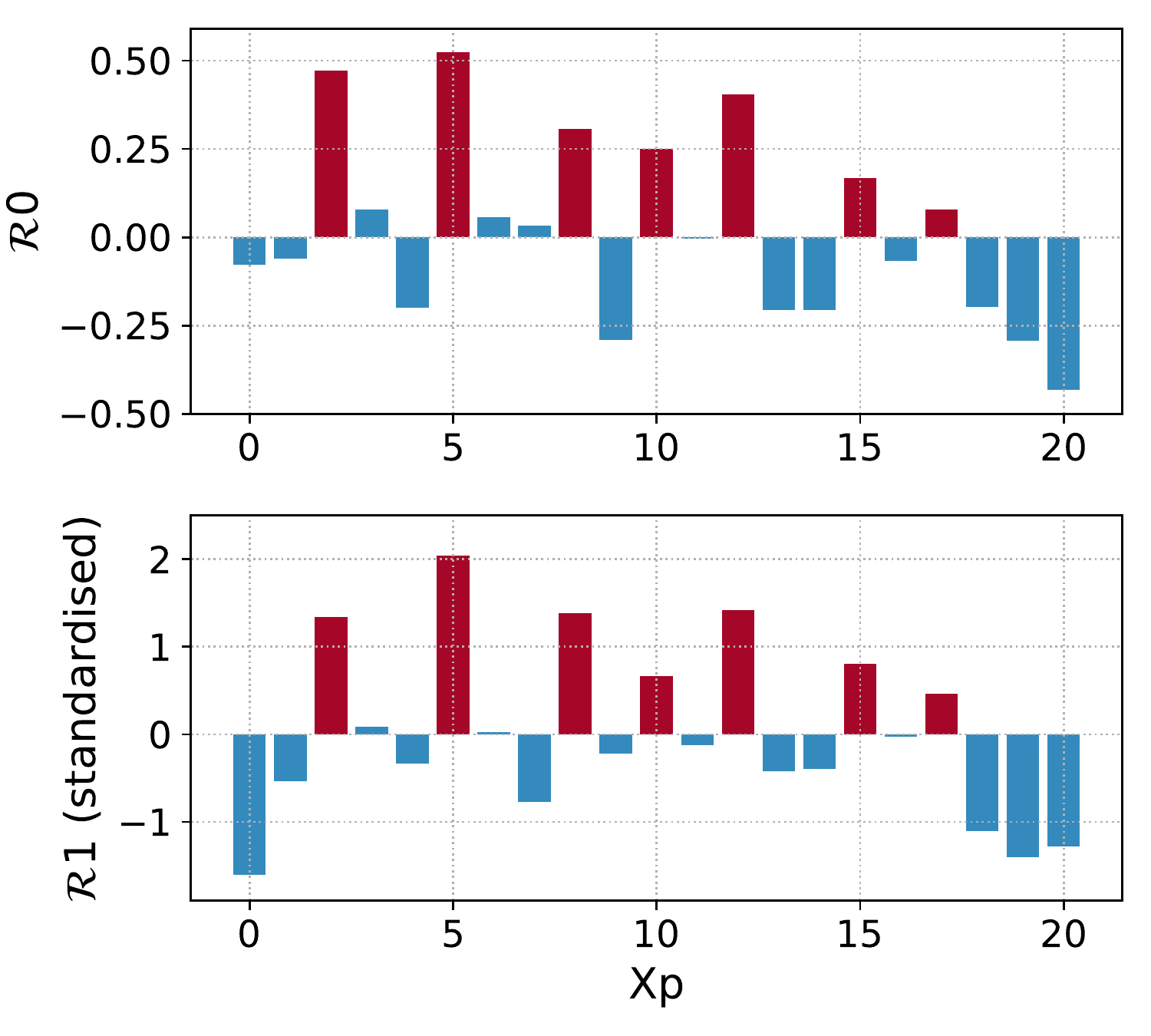}}
\centerline{\includegraphics[width=0.5\textwidth,angle=0]{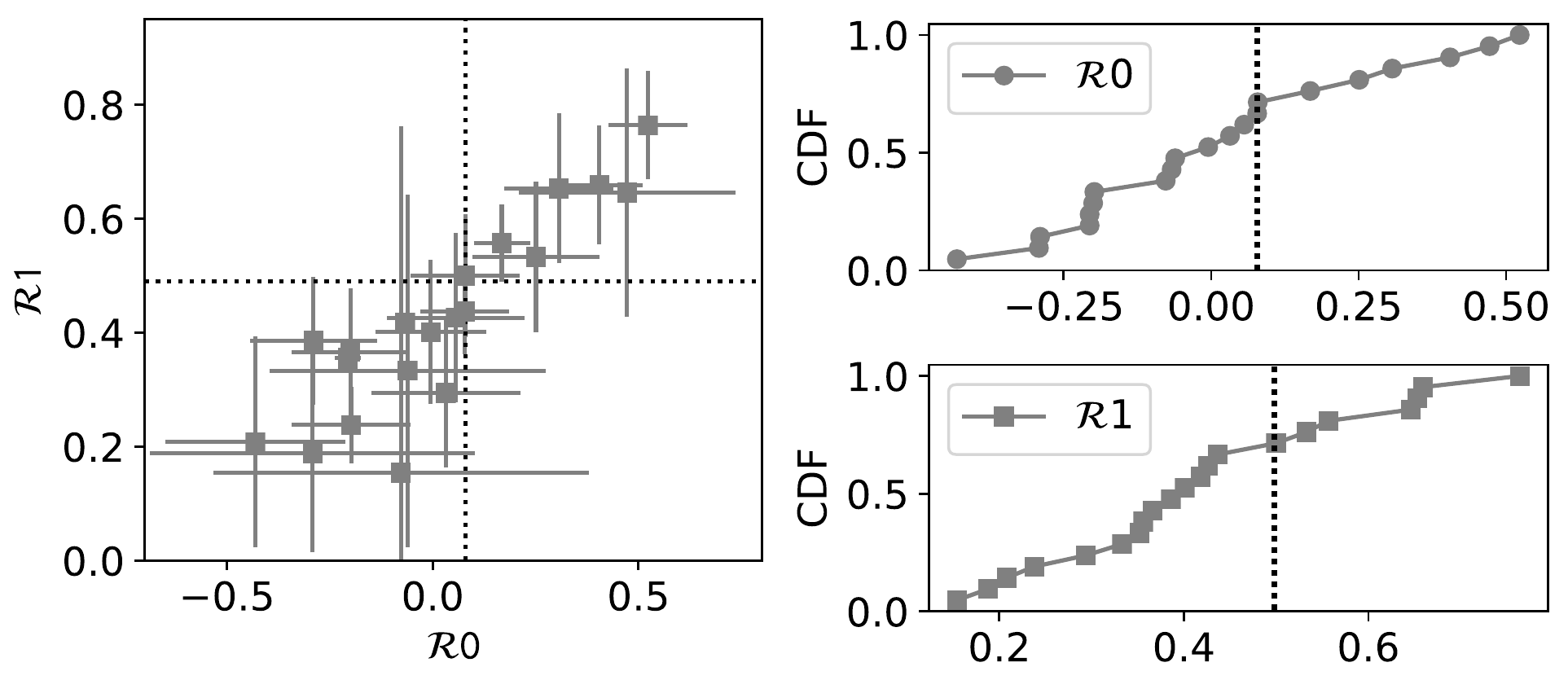}}
\caption{Two measures of Fe~{\sc xvv} excess, $\mathcal{R}0$ and $\mathcal{R}1$ (see text for their definition), for the 21 discrete sources, Xp0 through Xp20. $\mathcal{R}1$ values in the bar plot are standardised. Seven selected Fe {\sc xxv} (NB2)-excess source candidates are marked in red. Lower left: $\mathcal{R}0$ - $\mathcal{R}1$ diagram. Lower right: CDFs of $\mathcal{R}0$ and $\mathcal{R}1$. In both panels, dotted lines indicate the adopted thresholds for the NB2-excess selection.}
\label{fig:barplot}
\end{figure}

We use narrow-band imaging to select objects emitting strong high-ionization Fe K$\alpha $ emission from Fe {\sc xxv} at 6.7 keV among the 21 sources. The following three narrow bands, each of which has a 0.35 keV energy-interval, were set for imaging: NB0 (5.85-6.2 keV); NB1 (6.2-6.55 keV); and NB2 (6.55-6.9 keV). NB0 is the continuum band, while NB1 and NB2 might contain a cold Fe line at 6.4 keV and high-ionization Fe {\sc xxv} at 6.7 keV, respectively, if those lines are present. A higher energy band above NB2 was not used due to poor signal-to-noise ratio since the detector response declines rapidly above 7 keV. Our objective here is to find sources that shows a NB2 excess among the three bands. 

A measure of NB2 excess against the neighbouring NB1 band was defined as $\mathcal{R}0=({\rm NB2}-{\rm NB1})/({\rm NB1}+{\rm NB2})$. $\mathcal{R}0$, by definition, takes values between $-1$ and $+1$. A positive value suggests an excess in NB2. It should be noted that a negative value does not necessarily mean a presence of a cold Fe line. Since a source spectrum, in general, declines towards higher energies and so does the detector response, even a featureless continuum spectrum would result in a negative $\mathcal{R}0$. A secondary measure, $\mathcal{R}1 = {\rm NB2}/({\rm NB0}+{\rm NB2})$ using NB0 is also introduced. While NB0 is not adjacent to NB2, it is a line-free band, unlike NB1 which might contain a 6.4 keV Fe line, and generally has more counts than in NB1, making $\mathcal{R}1$ complementary to $\mathcal{R}0$. $\mathcal{R}1$ always takes a positive value between 0 and 1, and a source with a large $\mathcal{R}1$ is more likely to have a strong Fe {\sc xxv} line. 

Both $\mathcal{R}0$ and $\mathcal{R}1$ values averaged over each source region of the 21 discrete sources are computed and shown in Fig. \ref{fig:barplot}. For an easier inspection, the $\mathcal{R}1$ values are standardised (zero-centred, the standard deviation of 1) and shown in the bar-plot.
$\mathcal{R}0$ and $\mathcal{R}1$ are strongly correlated, as expected (Fig. \ref{fig:barplot}). As virtually no Fe~{\sc xxv} in X-ray binaries but strong Fe {\sc xxv} (EW $\ga 0.5$ keV) in SNRs are expected, the underlying distribution of $\mathcal{R}0$ and $\mathcal{R}1$ should be bimodal because line emitters with intermediate EW of 0.2 keV, for example, are unlikely to exist, but variations in continuum slope and line strength, as well as statistical noise, will add scatters that will blur the bimodality in the observed distribution. The cumulative distribution function (CDF) of $\mathcal{R}0$ and $\mathcal{R}1$ (Fig. \ref{fig:barplot}) suggests whereby a high-value excess in the distribution starts to arise. We therefore adopted the following thresholds: $\mathcal{R}0\geq0.08$ and $\mathcal{R}1\geq 0.5$, which roughly correspond to Fe~{\sc xxv} of EW$>0.3$ keV, to select Fe~{\sc xxv} (NB2)-excess sources. These selection criteria should be reasonably robust since a false positive rate is below 1.5\% for most sources (details of this examination using simulations are given in Appendix B).

The above criteria selected seven sources (Xp2, 5, 8, 10, 12, 15, and 17) as promising NB2-excess candidates.
Their position, 4-8 keV source counts, $\mathcal{R}0$ and $\mathcal{R}1$ values are given in Table \ref{tab:hotseven}. Among the selected seven, Xp2 is exceptionally faint and its spectrum, in fact, has small counts (five counts) in the Fe line. Therefore, we tentatively consider its selection to be insecure and Xp2 is excluded from further analysis (but see Sect. 3.6). We thus selected the remaining six sources as candidates of Fe~{\sc xxv} emitting X-ray SNRs. Our narrow-band imaging selection successfully recovered the Fe {\sc xxv} emitter identified by \citet{gandhi11} and \citet{liu14}, which is Xp5. The other X-ray SNR identified by \citet{kong07,chiang11} is, too, selected as Xp15, which now appears to be a promising Fe {\sc xxv} emitter.

The $\mathcal{R}0$ and $\mathcal{R}1$ values of 14 discarded sources shown in Fig. \ref{fig:barplot} are given in the Table \ref{tab:others}. Among them, Xp3, 6, 7, and 11 have ($\mathcal{R}0, \mathcal{R}1$) values close to but below the thresholds. These four sources are, however, all found to show transient flaring, which rules them out as SNRs. The 4-7 keV light curves of all the 21 sources, obtained from the 13 observations (Table \ref{tab:obslog}), and their variability characteristics are presented as supplemental material in Appendix D. 
The faintest sources (Xp0, 1, and 19) remain ambiguous as they have too small counts to examine the Fe K band.


\begin{table}
\caption{Seven X-ray SNR candidates.}
\label{tab:hotseven}
\centering
\begin{tabular}{rccrcc}
  \hline\hline
Xp & RA & Dec. & Counts & $\mathcal{R}0$ & $\mathcal{R}1$ \\
  \hline
  5 & 52.7 & 45.8 & 758/675 & 0.52 (0.10) & 0.76 (0.10)\\
  8 & 52.0 & 45.3 & 344/218 & 0.31 (0.13) & 0.65 (0.13) \\
  10 & 51.9 & 42.9 & 307/235 & 0.25 (0.15) & 0.53 (0.13) \\
  12 & 51.5 & 45.8 & 545/443 & 0.41 (0.11) & 0.66 (0.10) \\
  15 & 50.6 & 43.6 & 1654/1528 & 0.17 (0.07) & 0.56 (0.07)\\
  17 & 50.1 & 43.7 & 479/366 & 0.08 (0.13) & 0.50 (0.11) \\
  \hline
  2 & 53.8 & 47.4 & 118/47 & 0.47 (0.41) & 0.65 (0.27) \\
  \hline
\end{tabular}
\tablefoot{Source positions, R.A. and Dec., are offset in arcsec from $09^{\rm h}55^{\rm m}00^{\rm s}$ and $+69\degr 40\arcmin 00\arcsec$ in J2000. The column of 'Counts' shows 4-8 keV counts accumulated from each source region over the 11 exposures used for constructing the spectra shown in Fig. \ref{fig:fn6} and estimate of net source counts corrected for the diffuse background. Values of the Fe~{\sc xxv} excess indicators, $\mathcal{R}0$ and $\mathcal{R}1$, for each source are given with their errors in parenthesis. Selection of Xp2 is insecure due to small counts for the Fe line.}
\end{table}

\subsection{Mean spectra and Fe K emission of selected sources}

\begin{figure*}
  \centerline{\includegraphics[width=0.85\textwidth,angle=0]{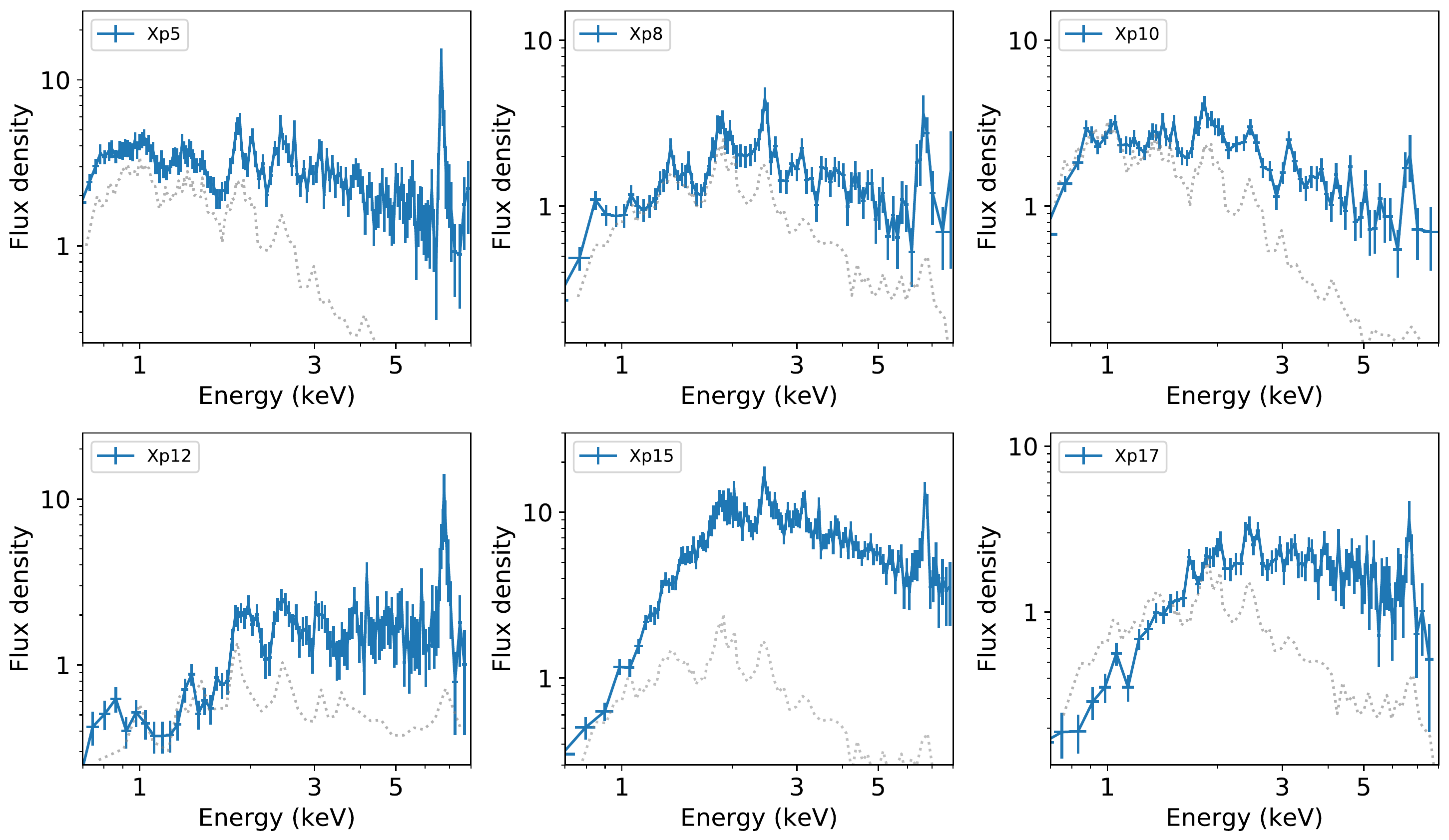}}
  \caption{Flux density spectra of the six selected Fe {\sc xxv}-emitting sources, obtained from the {\it Chandra} ACIS-S. Data were re-binned for display purpose. The flux density data are in units of $10^{-14}$ erg\thinspace cm$^{-2}$\thinspace s$^{-1}$\thinspace keV$^{-1}$. The estimated local diffuse background for each source is indicated by the dotted grey line. Fe {\sc xxv} lines at 6.7 keV are clearly detected in all the sources. The spectrum of Xp17 contains some contribution of the nearby bright source Xp16 spilling over into the Xp17 aperture.}
\label{fig:fn6}
\end{figure*}

To verify the presence of the Fe {\sc xxv} line, which is indicative of thermal emission, the mean spectra of the selected NB2-excess sources were constructed from the 11 exposures obtained from the ACIS-S3. The two ACIS-I exposures were discarded for the lower spectral resolution. The integrated exposure time for each source is 571 ks. These observations spanned 13 years over which the detector response below 2 keV has decreased and thus combining them is deemed to introduce uncertainty in the averaged response. However, as our spectral study is restricted to the 4-8 keV range, where a stable effective area was maintained, the uncertainty in the spectra is dominated by the diffuse background rather than the instrumental response, and thus we used the combined data. The spectral data were binned so that each bin contains at least one count and the $C$-statistic \citep{cash79} was used in spectral fitting. The best estimate of a spectral parameter was obtained by comparing between a spectral model folded through the detector response and the count-rate spectrum.

Figure \ref{fig:fn6} shows flux density spectra of the selected six objects. The spectral data were collected from the small apertures, ranging from $0\farcs 9$ to $1\farcs 1$, shown in Fig. \ref{fig:e48img} and include the foreground and background diffuse emission in which the sources are embedded. It is clear that all six spectra show strong Fe {\sc xxv} lines at $\sim 6.7$ keV as expected. Emission-lines of Si, S, Ar and Ca seen in lower energies, in excess above the background emission, in addition to Fe {\sc xxv}, support that the spectra are of thermal origin, suggesting that they are likely young X-ray SNRs.

The instrumental and cosmic backgrounds are negligibly small
due to the small apertures. Some sources are more absorbed than the others, resulted from varying obscuration across the region, which will be reported elsewhere. We approximated the diffuse background of each source by a spectrum taken from a neighbouring source-free region with an area 2 to 5 times larger than that of the source, correcting for aperture. The diffuse background estimated for each source is shown in their spectrum. The diffuse emission dominates below 2 keV in most sources. This is compatible with our argument that the majority of these sources are not visible at energies below 2 keV in the image. The diffuse background in Xp17 is overestimated below 1.5 keV, likely due to spatial fluctuation of absorption. While the diffuse emission spectrum also has a Fe {\sc xxv} line, the individual spectra clearly show much brighter Fe {\sc xxv} emission, asserting that the detected Fe lines originate primarily from the selected sources themselves. In fitting the 4-8 keV spectrum of each source, the diffuse background was modelled by a power-law with a Gaussian line for Fe {\sc xxv} and included as a fixed component. We note that the Xp17 spectrum contains significant flux from the nearby bright source Xp16, as argued in Appendix D. We estimated $\sim (3\pm 1)\times 10^{-14}$ \ergpspsqcm\ of the 4-8 keV continuum flux come from Xp16. The observed Fe {\sc xxv}, detected at 3-4 $\sigma $, is not affected since Xp16 does not show Fe emission. Xp15 may also be affected by the PSF tail of Xp16 but its contribution is less than 10\% of the Xp15 spectrum in 4-8 keV.

To characterise the spectral shape, we first fitted a power-law plus a Gaussian line for Fe {\sc xxv}. The power-law slope\footnote{We use the energy index, $\alpha$, which is apt for the spectra in units of flux density presented in this paper, instead of the conventionally used photon index, $\Gamma $, for photon-rate spectra. They are related as $\Gamma = 1 + \alpha$.}, $\alpha $, Fe line centroid energy, and EW of the line, obtained from the fit are reported in Table \ref{tab:specfit}. We also fitted the thermal emission spectrum computed by {\tt apec} for gas in a collisional ionization equilibrium (CIE). It has been known that X-ray emitting gas of SNRs are, in general, not in CIE \citep[e.g.][]{bleeker90,masai94}. However, a detailed study of non-CIE spectra is beyond the scope of this paper and we used the simpler CIE model to fit their temperatures. Given the limited 4-8 keV range, the temperature determination was driven by the bremsstrahlung continuum as well as the Fe K line.
Goodness of fit was assessed by comparing between the $C$-statistic for the observed spectrum and its expected value estimated by 2000 simulations based on the best-fit model. The $C$ values for all the fits but one fall within the 68\% compatible interval (CI) of the respective distribution of expected $C$ values, indicating that those fits are reasonably good. The only exception is the CIE fit for Xp12. Its $C$ value was found at the 96th percentile of the simulated $C$ distribution. This suboptimal fit was caused by a mismatch between the Fe K line and the model, which is looked into further in Sect. 3.5.2. The 4-8 keV fluxes, corresponding luminosities and the Fe K line luminosities were estimated and given in Table \ref{tab:specfit}. The uncertainties in flux and luminosity include estimated errors of the diffuse background due to its spatial variation. For Xp15 and Xp17, values affected by the contamination of Xp16 are corrected assuming the estimated contribution of Xp16. While the correction is small in Xp15 with the assumed $\sim 6$\% contamination, the uncertainties of those values for Xp17 become large, as they are dominated by the uncertainty of the contaminating flux estimate.


\begin{table*}
\caption{X-ray properties in the 4-8 keV band.}
\label{tab:specfit}
\centering
\begin{tabular}{rccccccc}
  \hline\hline
  Xp & $\alpha $ & $E_{\rm Fe}$ & EW & $kT$ & $f_{\rm 4-8}$ & $L_{\rm 4-8}$ & $L_{\rm Fe}$\\
   & & keV & keV & keV & & & \\
  \hline
  5 & $0.4^{+0.3}_{-0.3}$ & $6.66^{+0.01}_{-0.01}$ & $1.2^{+0.2}_{-0.2}$ & $5.4^{+0.7}_{-0.8}$ & 7.6\,(0.3) & 12\,(0.5) & 2.6\,(0.4) \\
  8 & $0.4^{+0.5}_{-0.6}$ & $6.70^{+0.02}_{-0.03}$ & $1.0^{+0.3}_{-0.3}$ & $6.3^{+1.9}_{-1.3}$ & 3.4\,(0.3) & 5.3\,(0.5) & 0.97\,(0.3) \\
  10 & $1.0^{+0.5}_{-0.5}$ & $6.62^{+0.04}_{-0.03}$ & $0.5^{+0.2}_{-0.2}$ & $5.5^{+2.6}_{-1.7}$ & 3.1\,(0.3) & 4.8\,(0.5) & 0.45\,(0.2)\\
  12 & $0.0^{+0.4}_{-0.3}$ & $6.74^{+0.02}_{-0.01}$ & $1.4^{+0.2}_{-0.3}$ & $7.4^{+1.1}_{-0.9}$ & 6.3\,(0.6) & 9.8\,(0.9) & 2.7\,(0.5)\\
  15 & $1.2^{+0.3}_{-0.2}$ & $6.71^{+0.02}_{-0.02}$ & $0.6^{+0.1}_{-0.1}$ & $5.0^{+0.6}_{-0.7}$ & 16\,(1) & 25\,(1.6) & 2.8\,(0.6)\\
  17$^{\star}$ & $1.1^{+0.4}_{-0.4}$ & $6.66^{+0.04}_{-0.04}$ & $1.0^{+0.6}_{-0.5}$ & $4.8^{+1.5}_{-0.8}$ & 2\,(1) & 3\,(1.5) & 0.61\,(0.2)\\
  \hline
\end{tabular}
\tablefoot{Energy index, $\alpha $, Fe K line energy, $E_{\rm Fe}$ and line equivalent width, EW, were obtained by fitting a power-law plus a Gaussian. Temperature $kT$ was estimated by fitting {\tt apec} model for CIE plasma. Mean source flux in the 4-8 keV band, $f_{\rm 4-8}$, is in units of $10^{-14}$ erg\thinspace cm$^{-2}$\thinspace s$^{-1}$\thinspace keV$^{-1}$ and is corrected for the diffuse background. The 4-8 keV and Fe line luminosities, $L_{\rm 4-8}$, $L_{\rm Fe}$, are given in units of $10^{37}$ \ergps. The line luminosity was derived from the Gaussian fit. Uncertainty of each flux or luminosity value is given in parenthesis. $^{\star}$The values of EW, $f_{4-8}$ and $L_{4-8}$ have been corrected for the contamination of Xp16. The Fe line detection is $>3\sigma $ but the large error of EW is due to the uncertainty of the contamination estimate.}
\end{table*}

\subsection{Xp12: X-ray counterpart of SN2008iz}

\subsubsection{Emergence of an X-ray source}

\begin{figure}
\centerline{\includegraphics[width=0.35\textwidth,angle=0]{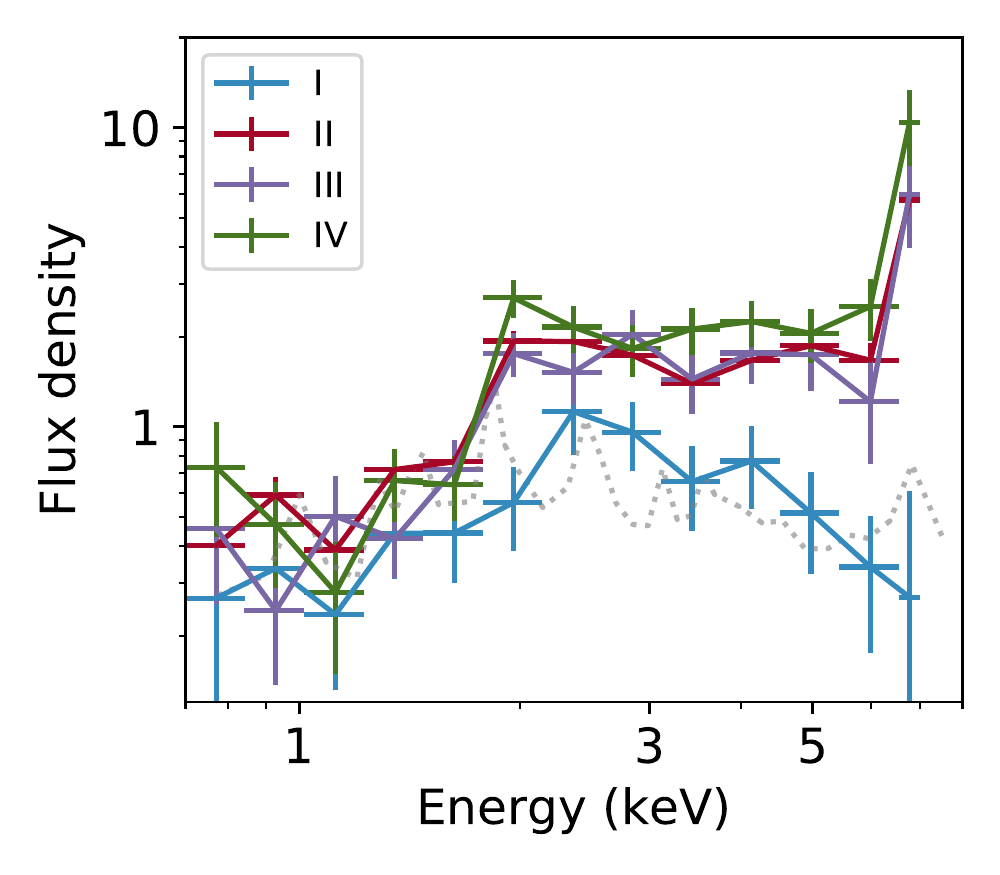}}
\caption{{\it Chandra} spectra of Xp12, taken in the four periods I (1999-2002), II (2009), III (2010), and IV (2012-2015); see Table \ref{tab:obslog}. Each 0.7-7 keV low-resolution spectrum has been binned with 12 logarithmically equal intervals over 0.7-6.55 keV and a 6.55-7 keV interval for the Fe {\sc xxv} emission at the highest energy end. The spectrum of Period I is comparable with the surrounding diffuse emission, indicated by dotted grey line.} 
\label{fig:xp12spvar}
\end{figure}

\begin{figure}
\centerline{\includegraphics[width=0.5\textwidth,angle=0]{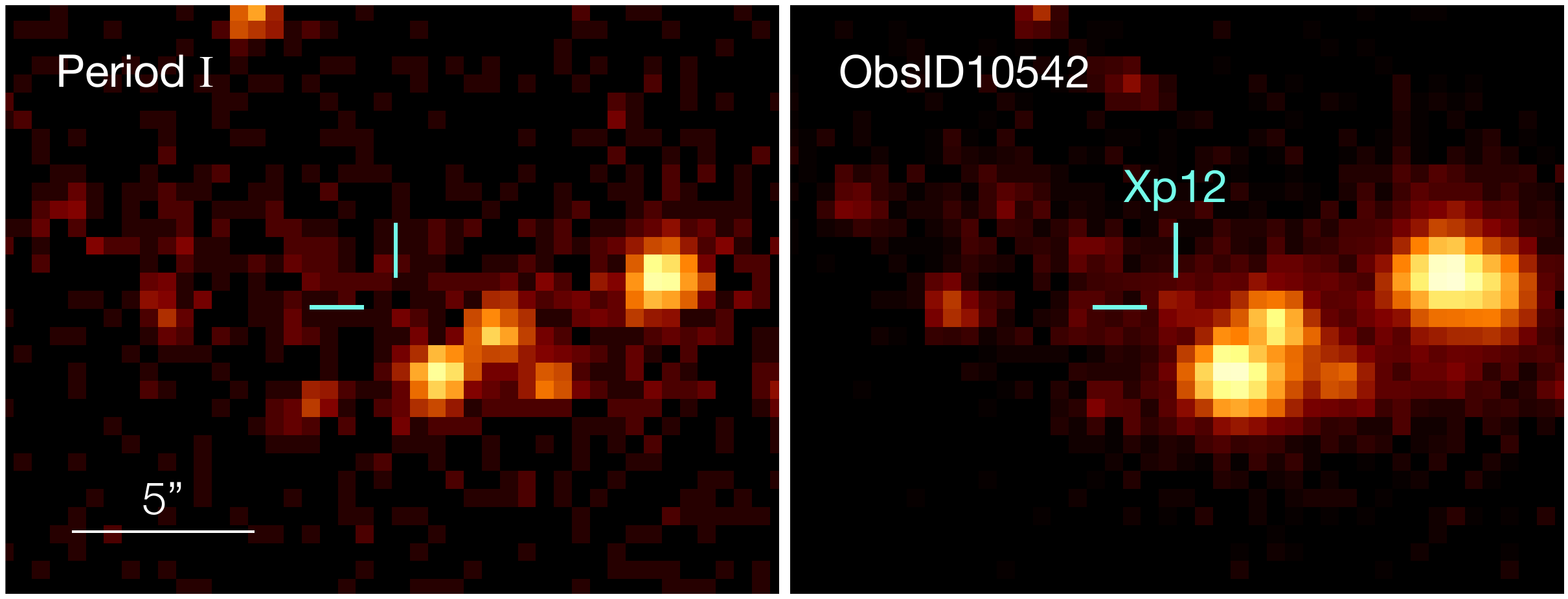}}
\caption{{\it Chandra} images of 5-7.5 keV centred on the position of Xp12 (as indicated by two lines) taken in the three observations in Period I (1999-2002) and ObsID~10542 in 2009. The exposure times are 66 ks and 118 ks, respectively. Observed counts in the $0\farcs 9$ aperture centred on Xp12 are 12 counts (the contribution of diffuse emission counts for the same aperture estimated from the surrounding region is $12\pm 3.5$ counts, meaning no excess) in Period I. In ObsID~10542, the image registered 56 counts for the Xp12 aperture, including the estimated diffuse emission of $25\pm 5$ counts, leaving an excess of 31 counts.}
\label{fig:xp12img}
\end{figure}

Xp12 exhibits an interesting X-ray variability. Figure \ref{fig:xp12spvar} shows the low-resolution 0.7-7 keV spectra of Xp12 taken from the four periods, I, II, III, and IV denoted in Table \ref{tab:obslog}. The spectrum of the earliest three observations in Period I is markedly softer than the other spectra. Its flux level is actually comparable to that of estimated diffuse background as indicated. The spectra from the following three periods are hard and show a strong Fe line, as seen in the mean spectrum in Fig. \ref{fig:fn6}. The source brightness is similar between Period II and III, which are only 1 yr apart, and is comparable or possibly elevated by $\sim 30$\% in Period IV of 2-5 yr later. This suggests X-ray emission from this Fe {\sc xxv} source has appeared in the observational gap in 2002-2009 between Periods I and II.

We turned to the {\it Chandra} imaging data for a further inspection. Figure \ref{fig:xp12img} shows a comparison of the {\it Chandra} images of the Xp12 region integrated over the three observations in Period I and that of ObsID~10542, the first observation of Period II. The images were constructed in the 5-7.5 keV band so that the contribution of the diffuse emission is reduced, compared to that in the 4-8 keV image. The exposure times are 66 ks for Period I and 118 ks for ObsID~10542. No excess emission above the surrounding diffuse emission is seen at the position of Xp12 in the image of Period I. On the other hand, excess counts of 31 that are $\sim 6$ times larger than the estimated background fluctuation is observed in the ObsID~10542. This result suggests that X-ray emission of Xp12 arose between the two periods.

We note that \citet{brunthaler09} reported a bright radio transient detected by VLA that appeared at the position of Xp12 in 2008. Follow-up observations show an expansion of the radio source \citep{brunthaler10,kimani16} supporting the SNR nature and the radio source is designated as SN2008iz. Its explosion at optical wavelengths was not recorded due to heavy obscuration: \citet{brunthaler10} estimated a column density towards the radio source to be \nH $= 5.4\times 10^{22}$\psqcm\ from the CO map of \citet[][but see \citealp{mattila13}]{weiss10}. Fitting an absorbed thermal emission spectrum to our 2-4 keV data gives a similar absorbing column density of \nH $=(7\pm 2)\times 10^{22}$ \psqcm\ towards Xp12.
\citet{brunthaler10} reported no {\it Chandra} detection, probably because of short exposures and large off-axis angles of the observations they used and soft X-ray suppression due to the absorption. 
The explosion date is estimated to be 2008 Feb 18 from the radio light curve \citep{marchili10}. This means that the {\it Chandra} observations in Period II captured the SN 16 months after its explosion.

\subsubsection{Fe K line profile}

\begin{figure}
\centerline{\includegraphics[width=0.5\textwidth,angle=0]{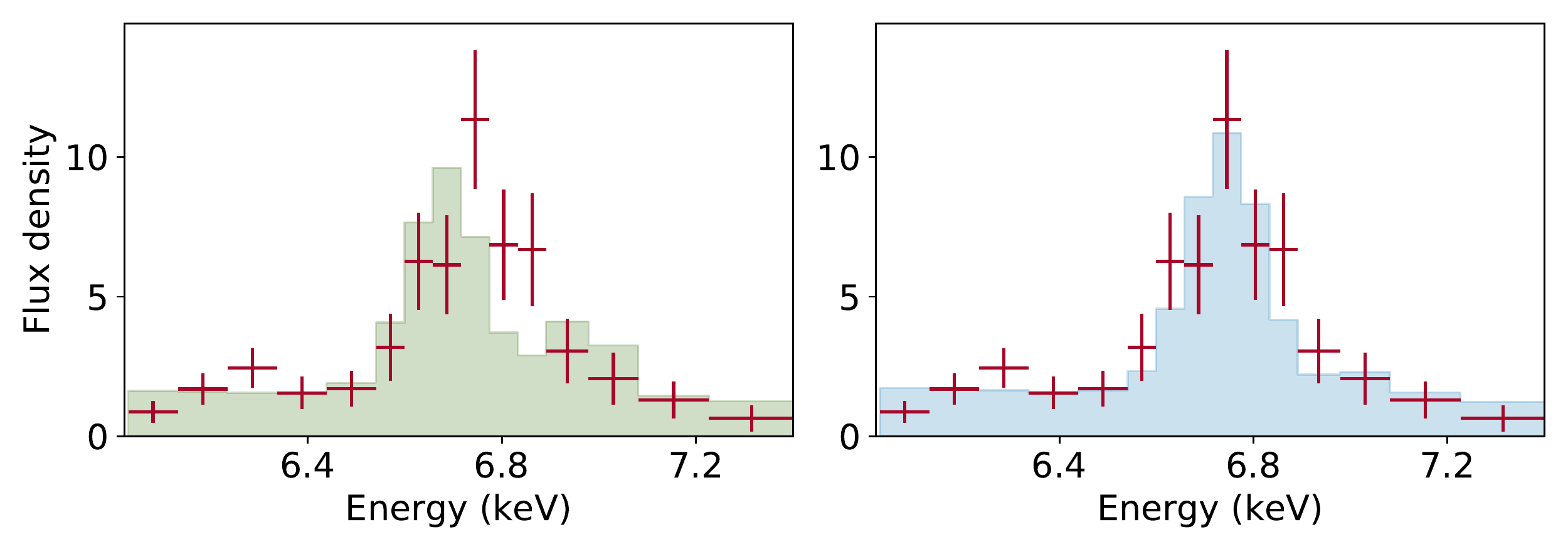}}
\caption{Left: Fe K band spectrum of Xp12 (red) with the best-fit {\tt apec} model with no velocity shift (green shade) relative to the systemic velocity (203 km\thinspace s$^{-1}$). Right: Same data but with the model with the velocity shift being fitted (blue shade, $v=-2700$ km\thinspace s$^{-1}$). }
\label{fig:xp12feline}
\end{figure}

Xp12 stands out from the other sources based on its hard spectrum and strong Fe K line (Table \ref{tab:specfit}). Another peculiarity is the centroid energy of the Fe K feature is well determined at $6.74^{+0.02}_{-0.01}$ keV, which is slightly higher than expected for Fe {\sc xxv}. The Fe {\sc xxv} emission is a triplet, composed of lines emitted at energies between 6.62 keV and 6.70 keV \citep{bautista00,oelgoetz04} that are blended together at the CCD resolution so that its centroid would fall at 6.7 keV or slightly below, depending on relative strengths of the triplet. This is indeed the case for the lines in the other sources (Table \ref{tab:specfit}). The centroid energy higher than 6.7 keV might therefore indicate a bluewards velocity shift. Fig. \ref{fig:xp12feline} shows the Fe K emission spectrum of Xp12. We tested whether the introduction of a velocity shift explains the data better. We used the limited energy range (6.0-7.4 keV) centred on the Fe K line so that the line feature, rather than the continuum, is the main driver in finding the best-fit solution for a CIE thermal emission spectrum.
In the interest of an optimised line profile study, we binned the spectral data, following the binning strategy discussed in \citet{kaastra16}. The Fe K line band 6.55-6.9 keV were binned at 60-eV intervals, which corresponds to $\sim 1/3 {\rm FWHM}$ in this band, so that the whole line profile is well over-sampled, resulting in an optimal sensitivity to a detectable velocity shift at the given resolution. The neighbouring faint continuum bands were binned at broader intervals. This ensures non-zero counts in each bin and stability of the simulations for testing goodness of fit to follow.

A fit assuming no velocity shift from the galaxy's systemic velocity \citep[$v=+203$ km\thinspace s$^{-1}$][]{goetz90,shopbell98} is displaced from the data to lower energies, leaving the blue side of the line profile unexplained (Fig. 6 Left). The best-fit temperature is found to be $kT = 6.9^{+1.3}_{-1.0}$ keV. When a velocity shift was fitted, the quality of fit improved (Fig 6 Right) and a blueshift of $z=-8.9^{+1.3}_{-3.0}\times 10^{-3}$ was obtained, corresponding to $v\simeq -2700^{+600}_{-500}$ km/s. The temperature is found to be $kT=4.6^{+1.4}_{-0.9}$ keV.

Since the fits used $C$-statistic, as defined as $C=-2ln\mathcal{L}$, where $\mathcal{L}$ is the likelihood for data of a Poisson distribution \citep{cash79}, the Akaike or Bayesian information criterion, AIC \citep{akaike74} or BIC \citep{schwarz78}, for each fit can be directly calculated.
The reduction of $C$, $\Delta C= -12.6$, when the extra parameter (velocity shift) was introduced, corresponds to $\Delta \mathrm{AIC}=-10.6$ and $\Delta \mathrm{BIC}=-9.9$, both of which indicate reasonably strong evidence for the velocity shift. For example, the model with no velocity shift is exp$(\Delta {\rm AIC}/2)\simeq 0.005$ times as probable as that with the blueshift.
Goodness of fit was again tested against the expected $C$ value through simulations. The fit of the model with the velocity shift gave $C=12.4$, close to the expected value $C=13$ and well within the 20\% CI, indicating a good fit. For the model with no velocity shift, expected $C$ is also, naturally, $C=13$, which deviates from $C=25.0$ obtained for the observed data, indicating no velocity-shift being less probable, as discussed above.  

Detection of the blueshift suggests a kinetically active state of the shocked medium just a few years after the SN explosion. It might also suggest an anisotropic expansion of the Fe emitting ejecta, as seen in Cas A \citep[e.g.][]{delaney10, hwang04}, otherwise symmetric line-broadening would be expected. Among extragalactic SNe, \citet{quirola19} reported a similar blueshift ($v\approx -2000$ km~s$^{-1}$) in Fe {\sc xxvi} of SN1996cr.

\subsection{Radio counterparts}

Radio imaging detected a number of discrete sources across the central molecular disc in M82 \citep[e.g.][]{kronberg85,huang94,muxlow94,pedlar99,mcdonald01,fenech08}. We compare the X-ray selected SNRs with the 32 radio SNRs detected by VLBI and MERLIN imaging at 1.7 GHz \citep{fenech10}, as shown in Fig. \ref{fig:e48img}. The four brightest radio sources coincide with the X-ray sources Xp2, Xp5, Xp8, and Xp15 (Table \ref{tab:radio}). All the four X-ray sources are those selected as Fe {\sc xxv} sources in Sect. 3.3 (Table \ref{tab:hotseven}), although the SNR nature of Xp2 was considered to be uncertain due to marginal line detection. Its correspondence to the radio SNR supports Xp2 to be also a possible X-ray SNR (the X-ray spectrum and other X-ray properties are given in Appendix E). These four radio SNRs are all compact with the diameters smaller than 1.2 pc. The brightest and most compact radio SNR among those reported in \citet{fenech10} is 41.95+57.5, identified with Xp15, also the brightest in X-ray. This radio SNR is peculiar as it has a double-lobed morphology rather than shell-like usually seen in SNRs.

Previously, possible associations between the X-ray and radio sources have been examined \citep{matsumoto01,strickland07,kong07,chiang11}.
Those works used various combinations of {\it Chandra} observations, including HRC observations, taken before 2007, for which the effective exposures in the 4-8 keV band are $\ga 9$ times shorter than that in this work.  
Although \citet{strickland07} concluded that the radio sources are generally not associated with X-ray point sources, they paired the brightest Xp15 and Xp5 with the radio sources.

As discussed above, Xp12 is the X-ray counterpart of the radio supernova SN2008iz (Table \ref{tab:radio}) and is consequently absent in \citet{fenech10} as their observations were carried out in 2005. Xp10 and Xp17 have no radio counterparts. These two are fainter X-ray sources but still brighter than Xp2.

The previous surveys of extragalactic SNRs in nearby galaxies found various degrees of overlap between SNRs selected at optical, radio, and X-ray \citep{pannuti07,long10,leonidaki10}, in which selection effects inherent to these searches at different wavelengths likely play a significant role. In M82, although radio SNR candidates outnumber the X-ray SNRs, it resulted from more sensitive surveys in radio than in X-ray. The five brightest radio SNRs -- four from the \citet{fenech10} sample and SN2008iz, which was the brightest radio source in M82 when discovered \citep{brunthaler09} -- are among the seven brightest X-ray SNRs, indicating a large overlap between the radio and X-ray SNRs at the bright end, relative to those found in the previous work mentioned above. This may be related to the dense environment of the starburst region they reside, which provides a favourable condition for a SNR to be radiative at both wavelengths \cite[e.g.][]{chevalier01}. It is also interesting to note that the brightest X-ray sources (Xp5 and 15) coincide with the radio sources of the expansion velocity markedly lower ($\simeq 2700$ km~s$^{-1}$ and $\simeq 1500$ km~s$^{-1}$, respectively, \citet{fenech08}), compared to the others with $\sim 10000$-20000 km~s$^{-1}$ \citep{mcdonald02,fenech08,beswick06,brunthaler10,kimani16}.

There are a few other radio SNRs in \citet{fenech10} that lie within the positional uncertainty of the X-ray sources: 44.40+61.8 and 45.24+65.2 near Xp3; and 43.72+62.0 at Xp6. However, both Xp3 and Xp6 show transient, strong brightening (see Appendix D), suggesting their primary sources are X-ray binaries. Although this does not readily rule out a contribution of the SNRs to the X-ray emission in their quiescence, no strong evidence of excess Fe~{\sc xxv} was found in the quiescence spectra of small (100-120) counts. 


\begin{table}
\caption{Radio counterparts.}
\label{tab:radio}
\centering
\begin{tabular}{lccc}
  \hline\hline
Xp & Name$_R$ & $F_{1.7}$ & Diameter \\
&& mJy & pc\\
\hline
2 & 45.17+61.2 & 17.60\,(0.16) & $1.2\times 1.0$ \\ 
5 & 44.01+59.6 & 12.32\,(0.12) & $1.0\times 0.8$ \\
8 & 43.31+59.2 & 23.55\,(0.35) & $0.9\times 0.8$ \\
12\tablefootmark{a} & 42.81+59.6 & 30\,(13) & $\approx 0.1$\\
15 & 41.95+57.5 & 38.25\,(0.21) & $0.4\times 0.3$ \\
  \hline
\end{tabular}
\tablefoot{The radio source names, their source diameters and fluxes were taken from Table 5 of \citet{fenech10}. Name$_R$ follows the naming convention of \citet{kronberg85} with the J1950 coordinates. The source diameters are corrected for our adopted distance of M82. The source flux density, $F_{1.7}$, is the total flux density at 1.7 GHz in units of mJy with uncertainty in parenthesis.\\
  \tablefoottext{a}{The radio counterpart of Xp12 was not observed by \citet{fenech10} as the SN explosion (SN2008iz) occurred after their observation. The quoted values are flux density at 1.6 GHz measured by VLBI and the diameter measured in August 2010, 1000 days after the explosion \citep[][the diameter was derived from their $R_{50}$]{kimani16}.}}
\end{table}

\subsection{X-ray variability of SNR candidates}

\begin{figure}
\centerline{\includegraphics[width=0.5\textwidth,angle=0]{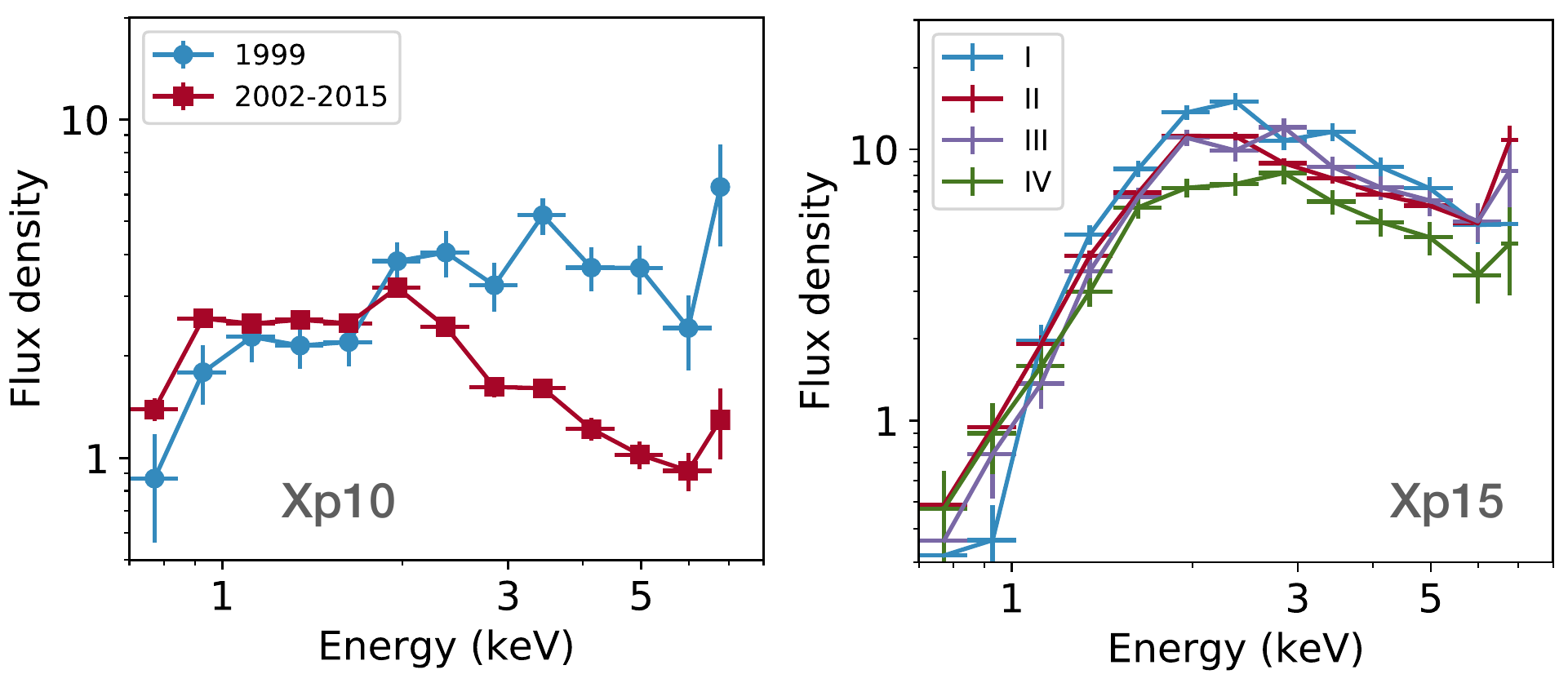}}
\caption{Low-resolution 0.7-7 keV spectra of Xp10 (Left) and Xp15 (Right), taken in various epochs. For Xp10, data from the two ACIS-I observations in 1999 and all the other 11 exposures (the same data as presented in Fig. \ref{fig:fn6}) are shown. For Xp15, data from the four periods between 1999 and 2015 (Table \ref{tab:obslog}) are shown. The flux density units are the same as in Fig. \ref{fig:fn6}.} 
\label{fig:xp15spvar}
\end{figure}

An inspection of light curves of the SNR candidates shows that, Xp10 and Xp15 appear to show variability (Table \ref{tab:var}) in addition to Xp12 discussed in Sect. 3.5. Xp10 was $\sim 3$ times brighter in the first two exposures in 1999 than in the rest of the observations in 2002 onwards. As the 1999 observations were performed with the ACIS-I, they are not included in the spectrum of Xp10 shown in Fig. \ref{fig:fn6}. Whilst the bright state could result from an unrelated transient source, an alternative explanation is that the 1999 observations caught an epoch relatively early, say $\la 1$ yr, after a SN explosion, since the spectrum shows a strong high-ionization Fe K excess at $6.95^{+0.4}_{-0.25}$ keV of Fe {\sc xxvi} with EW $= 1.7^{+2.5}_{-0.5}$ keV on a hard continuum of $\alpha = 0.5\pm 1.0$ (Fig. \ref{fig:xp15spvar}). The spectrum is well described by the thermal emission model {\tt apec} with a temperature of $kT = 10^{+20}_{-5}$ keV,
compatible with X-ray emission of a SN in its earliest phase \citep[e.g.][]{fox00}. The 2-8 keV luminosity is estimated to be $\simeq 4\times 10^{38}$ \ergps, which is comparable or slightly less luminous than SN1993J of a few months to 1 yr old \citep{chandra09}. Then the observed luminosity decline by a factor of $\sim 3$ in 3 yr is within expectations. If this is true, it would make Xp10 the second youngest SNR after Xp12.

The brightest source, Xp15, shows a decreasing trend in flux by $\sim 45$\% over 16 yr, as shown in Fig. \ref{fig:xp15spvar} (also the light curve in Appendix D). The radio counterpart 41.95+57.5 has been reported to show a persistent flux decay of 8.5\% per year \citep{fenech10}, which is slightly faster than but comparable to that in X-ray. It was first detected in 1968 \citep{bash68}, which means its age is $>40 $ yr old in 2008. Despite the age, it maintains the exceptionally large X-ray luminosity among the selected X-ray SNRs and is more luminous than the youngest Xp12.

On the other hand, Xp12 possibly shows X-ray brightness increasing with time since the first X-ray detection, as illustrated in the light curve (Fig. \ref{fig:lc}, also Fig. \ref{fig:xp12spvar}), which is unusual but have been seen in a small number of SNe such as SN1987A and SN1996cr \citep{quirola19,dwarkadas14}. Likewise, the flux of the relatively bright Xp5 is rather stable over 16 yr without a sign of declining.

\section{Discussion}

\subsection{Supernova rate}

One new X-ray SN (SN2008iz) appeared in the starburst region of M82 over the 16 years of {\it Chandra} observations we used (SN2004am, which lies outside the region was excluded). This is in agreement with what is expected for SFR $\sim 10$ \Msunpyr, but the SN rate inferred by observing one SN in 16 yr has a rather loose constraint: Assuming SNe follow the Poisson distribution, with a flat prior, this gives a SN rate of $\nu_\mathrm{SN} = 0.06$ yr$^{-1}$ with the 89\% CI of (0.02-0.28) yr$^{-1}$. 

Age constrains on the X-ray SNRs could give a further constraint.
Cas~A, one of the youngest known Galactic SNRs, is $\sim 350$ yr old and its 4-8 keV luminosity is $\sim 5\times 10^{35}$ \ergps, as estimated by our reanalysis of the ASCA GIS data. The selected X-ray SNRs in M82 all emit at $>10^{37}$ \ergps in the same band (Table \ref{tab:specfit}), suggesting that they are much younger than Cas~A.
Xp8 is the third least X-ray-luminous object among the six X-ray SNRs in M82. The radio counterpart of Xp8 (43.31+59.2) was first detected in 1972 \citep{kronberg75}, and \citet{fenech10} estimated its birth year to be 1952. Assuming more luminous SNRs are younger, the three SNe that became Xp8 and the more luminous Xp5 and Xp15 would have occurred between 1952 and 2008 until SN2008iz exploded. This yields the same SN rate of $\nu_\mathrm{SN}\approx 0.06$ yr$^{-1}$ with the 89\% CI of (0.03-0.13) yr$^{-1}$ from the updated posterior. This corresponds to a SFR of $\sim 10$ (5-17) \Msunpyr, assuming the Salpeter IMF. The SN rate estimated above is comparable to those derived from radio observations using various methods \citep[0.03-0.11 yr$^{-1}$;][]{kronberg75,rieke80,unger84,kronbergSramek85,muxlow94,huang94,fenech08,fenech10}.

Although the above supernova rate seems reasonable, the assumption on which the estimate is based may be too simplistic. It is generally true that X-ray light curves of extragalactic SNe show a continuous decline after the initial brightening, with a small number of exceptions \citep{dwarkadas14,ross17}, for example, the well-sampled 2-8 keV light curve of SN1993J shows a decline as $t^{-1}$ \citep{chandra09}. However, they are all optically identified unobscured SNe; whether this trend holds for our SNRs in the obscured environment is not entirely clear, since, as mentioned in Sect. 3.7, Xp12 may have been brightened for 7 yr since the SN explosion and Xp5 shows no sign of flux decline over 16 yr. Another issue is that the $L_\mathrm{X}$ of SNe at a given age can vary by a few orders of magnitude between SNe, especially at the early phase \citep{dwarkadas14}. Assuming our speculation in Sect. 3.7 is correct, Xp10 would be younger than Xp8, despite its $L_{4-8}$ being smaller than that of Xp8. In contrast, the most luminous Xp15 is estimated to be $\sim 80$ yr old (in 2008, \citealp{fenech08}). Given this possible diversity in the evolution of X-ray luminosity, the above estimate of the SN rate needs to be treated with caution. 

Nevertheless, the selected SNRs seem to generally belong to the luminous class. Compared to SN1993J, the $L_{2-8}$ of which dropped below $10^{37}$ \ergps\ after 20 yr \citep{kundu19}, all our X-ray SNRs have $L_{4-8}>10^{37}$ \ergps, despite some of their ages, which are known to be $>20$ yr from radio observations. Since our X-ray SNRs are all embedded in the starbursting molecular disc, the surrounding dense interstellar medium (ISM) could amplify X-ray brightness, as suggested for the X-ray-luminous class of Type IIn SNe \citep{fox00,ofek13,chandra15,ross17,quirola19}.

\subsection{Contribution of SNRs to the total Fe K emission}

The hot, diffuse emission (Fig. \ref{fig:e48img}) is of a great interest in the context of the superwind, as formulated by \citet{chevalier85}. An extensive discussion using the {\it Chandra} data is presented in \citet{strickland07,strickland09}. They estimated the total Fe {\sc xxv} line luminosity of the diffuse emission to be $L_{\rm Fe} = $(1.1-1.7)$\times 10^{38}$ \ergps, as observed from the central region of M82, which requires the mass loading factor ($\beta $) of the wind to be larger than unity, $\beta =$ 1-2.8. However, we now know that part of the above $L_{\rm Fe}$ comes from the X-ray SNRs described in this work. We evaluate the contribution of SNRs and the net line-luminosity of the diffuse emission below.

We constructed the diffuse emission spectrum from the ellipse region in Fig. \ref{fig:e48img} from the 11 ACIS-S exposures, excluding the regions for the 21 discrete sources. The measured Fe {\sc xxv} line luminosity is $0.8\times 10^{38}$ \ergps. When corrected for the areas taken away by the regions for the 21 sources assuming the mean emissivity over the areas, we obtained the line luminosity of $L_{\rm Fe} = 0.9\times 10^{38}$ \ergps\ for the diffuse emission enclosed in the whole ellipse. On the other hand, the summed Fe {\sc xxv} line luminosity coming from the six X-ray SNRs is $1.0\times 10^{38}$ \ergps, which is comparable to that emitted by the diffuse emission. Strictly speaking, while the SNR $L_{\rm Fe}$ has been corrected for the PSF wing that spills out from the source apertures ($\sim 10$\% for the typical $1\arcsec $ radius aperture; MARX documentation\footnote{https://space.mit.edu/cxc/marx/tests/PSF.html}), the diffuse $L_{\rm Fe}$ double-counted the spillage from the SNRs which amounts to $\approx 0.1\times 10^{38}$ \ergps. Correcting for that, we estimated the diffuse $L_{\rm Fe}$ to be $(0.8\pm 0.1) \times 10^{38}$ \ergps. A sum of the line luminosities from the diffuse emission and SNRs agrees with the estimate by \citet{strickland07}, when taking into account that Xp12 (=SN2008iz), which is a luminous Fe {\sc xxv} source (Table \ref{tab:specfit}), was absent in the data they used. As a result, the requirement of highly mass-loaded winds could be relaxed.
It should also be noted that, to create hot ($T\sim 10^8$ K) fluid, hypothesised in the superwind model of \citet{chevalier85}, a low-density ISM needs to be heated directly by SNe. The X-ray SNRs we selected would not be efficient for that, because they lose substantial energy to radiation.

Similarly to the case of M82, SNRs could be an important source of Fe {\sc xxv} lines observed in starburst galaxies with similar or higher SFRs. An off-nuclear Fe {\sc xxv} clump found in NGC~4945 \citep{marinucci17}, for instance, might be a SNR. Some luminous or ultra-luminous infrared galaxies, such as Arp~220 and NGC~3690E, show strong Fe~{\sc xxv} in their spectra \citep{iwasawa05,iwasawa09,ballo04}, and their SNR origin was discussed in \citet{iwasawa05}. Since SNe in those galaxies are $>10$ times more frequent than in M82, Fe {\sc xxv} emission from SNRs would be accumulated efficiently and could be observed to be more luminous.

\begin{acknowledgements}
This research made use of data obtained from {\it Chandra X-ray Observatory} and the data archive maintained by {\it Chandra} X-ray Center (CXC). Software packages of CIAO~4.12, HEASoft~6.27.1, and IPython were used for data analysis. KI acknowledges support by the Spanish MICINN under grant Proyecto/AEI/10.13039/501100011033 and ``Unit of excellence Mar\'ia de Maeztu 2020-2023'' awarded to ICCUB (CEX2019-000918-M). 
\end{acknowledgements}

\bibliographystyle{aa} \bibliography{m82sn}{}

\begin{appendix}

\section{{\it Chandra} source names}
Official X-ray source names of the 21 discrete X-ray sources, Xp0 through Xp20, investigated in this work are listed in Table \ref{tab:cxonames}. Apart from four sources, all the others are registered in the {\it Chandra} Source Catalog 2.0 \citep[CSC~2.0][]{evans20}. These sources (2CXO) are listed as they appear in CSC~2.0. The unregistered sources (CXOU) are named, following the naming convention instructed in the guideline of the observatory. Some of the brightest sources (e.g. M82 X-1 and X-2) are ULXs that have been studied extensively. Eleven sources, including the brightests, have been detected by \citet{chiang11} and their source numbers are also given, as well as the brightest four sources as labelled in \citet{brightman19}. Accumulated 4-8 keV counts in the source aperture over the total 13 exposures are listed. We note that the source counts and other properties of the brightest three X-ray binaries (Xp13, 14, and 16) given here are all affected by photon pile-up. 

\begin{table}
  \caption{{\it Chandra} source names}
\label{tab:cxonames}
\centering
\begin{tabular}{rcccr}
  \hline\hline
  Xp & Name & CK11 & Other & Cts \\
  \hline
0 & 2CXO J095555.0+694054 & & & 67 \\
1 & 2CXO J095554.7+694053 & & & 71 \\
2 & CXOU J095553.8+694047 & & & 125 \\
3 & 2CXO J095553.8+694050 & 40 & & 741\\
4 & 2CXO J095553.1+694048 & 37 & & 412 \\
5 & 2CXO J095552.7+694045 & 36 & & 794 \\
6 & 2CXO J095552.4+694048 & & & 354 \\
7 & CXOU J095552.3+694053 & 34 & & 389\\
8 & 2CXO J095552.0+694045 & & & 370 \\
9 & CXOU J095552.0+694045 & & & 387\\
10 & CXOU J095551.9+694042 & 33 & & 404  \\
11 & 2CXO J095551.8+694051 & & & 516\\
12 & 2CXO J095551.5+694045 & & & 564\\
13 & 2CXO J095551.2+694043 & 30 & X-3 & 27291\\
14 & 2CXO J095550.9+694045 & 29 & X-2 & 9331\\
15 & 2CXO J095550.7+694043 & 28 & X-4 & 1845 \\
16 & 2CXO J095550.1+694046 & 25 & X-1 & 41010\\
17 & 2CXO J095550.1+694043 & & & 541\\
18 & 2CXO J095549.4+694043 & 23 & & 914\\
19 & 2CXO J095548.8+694039 & & & 83\\
20 & 2CXO J095548.8+694043 & 22 & & 181\\
\hline
\end{tabular}
\tablefoot{Name: official {\it Chandra} source name; CK11: Source numbers in \citet{chiang11} when available; Other: other X-ray source names (from \citealp{brightman19}); and Cts: 4-8 keV counts collected from each source aperture over the total 13 exposures. The counts of the brightest three sources (Xp13, 14, 16) are given as observed with no pile-up correction.}
\end{table}

\section{Selection reliability by $\mathcal{R}0$ and $\mathcal{R}1$}

\begin{figure}
\centerline{\includegraphics[width=0.5\textwidth,angle=0]{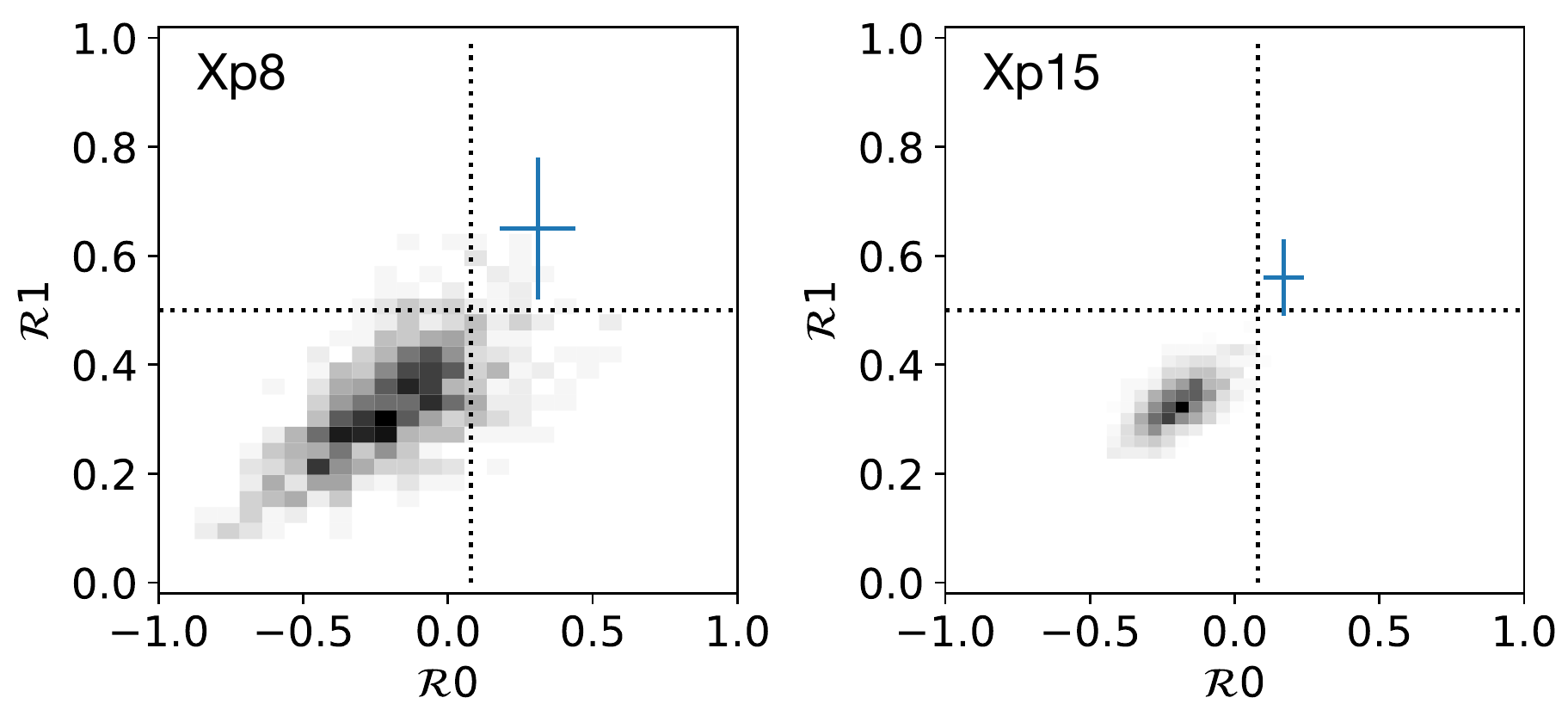}}
\caption{Bi-dimensional histograms of simulated $\mathcal{R}0$ and $\mathcal{R}1$, assuming a featureless power-law spectrum with the observed source brightness and spectral slope for Xp8 (Left) and Xp15 (Right). The vertical and horizontal dotted lines indicate the adopted thresholds for Fe {\sc xxv}-excess source selection, $\mathcal{R}0>0.08$ and $\mathcal{R}1 > 0.5$. The cross in each panel indicates the observed point.} 
\label{fig:sim}
\end{figure}

Upon inspection of their spectra, all the Fe {\sc xxv} (NB2)-excess sources selected through $\mathcal{R}0$ and $\mathcal{R}1$ based on the narrow-band imaging (Sect. 3.3) turned out to exhibit strong Fe {\sc xxv} lines at $\sim 6.7$ keV with EW = 0.5-1.2 keV (Sect. 3.4), though there was one marginal case (Xp2) due to low counts. We here demonstrate that the adopted selection thresholds for $\mathcal{R}0$ and $\mathcal{R}1$ yield sufficiently small false positives (i.e. a source with no Fe {\sc xxv} line is incorrectly selected) for the majority of the sources using simulations.

For each of the 21 detected sources, expected counts in the NB0, NB1 and NB2 bands in the case of a featureless spectrum were estimated by extrapolating a power-law continuum with the slope and brightness observed in the 4-6 keV band, given the exposure time. Assuming those expected counts follow the Poisson distribution, 1000 simulations for those three bands were performed and resulted $\mathcal{R}0$ and $\mathcal{R}1$ were computed. They represent the distribution of $\mathcal{R}0$ and $\mathcal{R}1$ when no Fe xxv line is present in the spectrum of an individual source. By applying the adopted selection thresholds ($\mathcal{R}0 > 0.08$ and $\mathcal{R}1 >0.5$) to the distribution, a false positive rate for given source can be estimated. This was repeated for all the 21 sources. Two examples for Xp8 and Xp15 are shown in Fig. \ref{fig:sim}. Both are the selected X-ray SNRs: Xp8 is the faintest but shows a strong (EW$\simeq 1$ keV) Fe {\sc xxv} line while Xp15 is the brightest with a line of moderate strength (EW$\simeq 0.5$ keV). The observed ($\mathcal{R}0$,$\mathcal{R}1$) are well distinct from the distributions assuming no Fe {\sc xxv} line. For the majority of the sources, the false positive rate is $<1.5$\%, or slightly higher (3\% for Xp20 and 5\% for Xp2). Only for the faintest three sources (Xp0, 1 and 19), $\mathcal{R}0$ and $\mathcal{R}1$ can be anywhere and thus have no constraining power. In summary, the adopted $\mathcal{R}0$ and $\mathcal{R}1$ selection criteria are expected to give a reasonably robust selection of Fe {\sc xxv}-excess sources for most sources, in agreement with the verification by the spectral analysis. 

\section{Discarded X-ray sources}
The 14 sources were filtered out as X-ray SNR candidates by the narrow-band imaging selection in Sect. 3.3. Their $\mathcal{R}0$ and $\mathcal{R}1$ values plotted in Fig. \ref{fig:barplot} are reproduced in Table \ref{tab:others}. The constraint (the 89\% upper limit) on the EW of a narrow line at 6.7 keV with respect to the continuum, which is composed of source emission and the diffuse background, for each source is also given. The 6.7 keV line detected in Xp11, one of the flaring sources (Appendix D), originates from the diffuse background, since there is no source excess in the image of the quiescent intervals. It is noted that the diffuse background emission has enhanced Fe {\sc xxv} in the region where Xp6, 7 and 11 reside. 

\begin{table}
\caption{Sources not selected as SNRs}
\label{tab:others}
\centering
\begin{tabular}{rccccc}
  \hline\hline
  Xp & RA & Dec & $\mathcal{R}0$ & $\mathcal{R}1$ & EW(Fe{\sc xxv}) \\
  \hline
  0 & 55.1 & 40.5 & $-0.08(0.45)$ & 0.15(0.61) & $<1.2$ \\
  1 & 54.7 & 53.7 & $-0.06(0.33)$ & 0.33(0.31) & $<1.9$ \\
  3 & 53.8 & 50.3 & $0.08(0.11)$ & 0.44(0.07) & $<0.20$ \\
  4 & 53.3 & 48.5 & $-0.20(0.14)$ & 0.37(0.11) & $<0.12$\\
  6 & 52.5 & 48.9 & $0.05(0.16)$ & 0.43(0.15) & $<0.45$\\
  7 & 52.3 & 53.7 & $0.03(0.18)$ & 0.29(0.13) & $<0.8$ \\
  9 & 52.0 & 45.3 & $-0.29(0.15)$ & 0.39(0.11) & $<0.20$ \\
  11 & 51.8 & 51.7 & $-0.01(0.13)$ & 0.40(0.13) & $0.45^{+0.16\star}_{-0.16}$\\
  13 & 51.3 & 43.6 & $-0.21(0.02)$ & 0.35(0.01) & $<0.01$ \\
  14 & 51.0 & 45.4 & $-0.21(0.03)$ & 0.36(0.02) & $<0.02$ \\
  16 & 50.1 & 46.5 & $-0.07(0.01)$ & 0.42(0.01) & $<0.01$ \\
  18 & 49.4 & 43.6 & $-0.20(0.14)$ & 0.24(0.06) & $<0.06$ \\
  19 & 48.9 & 39.8 & $-0.29(0.40)$ & 0.19(0.17) & $<0.24$\\
  20 & 48.9 & 43.7 & $-0.43(0.22)$ & 0.21(0.18) & $<0.26$\\
  \hline
\end{tabular}
\tablefoot{Source positions are offsets in arcsec from RA $= 09^{\rm h}55^{\rm m}00^{\rm s}$ and Dec. $= +69\degr 40\arcmin 00\arcsec$ (J2000). $\mathcal{R}0$ and $\mathcal{R}1$ are the Fe {\sc xvv}-excess indicators defined in Sect. 3.3. EW(Fe{\sc xxv}) is the equivalent width of a narrow 6.7 keV line in units of keV (or their 89\% upper limit). $^{\star}$This detection is from the background diffuse emission.}
\end{table}

\section{X-ray flux variability}

\begin{figure*}
\centerline{\includegraphics[width=0.98\textwidth,angle=0]{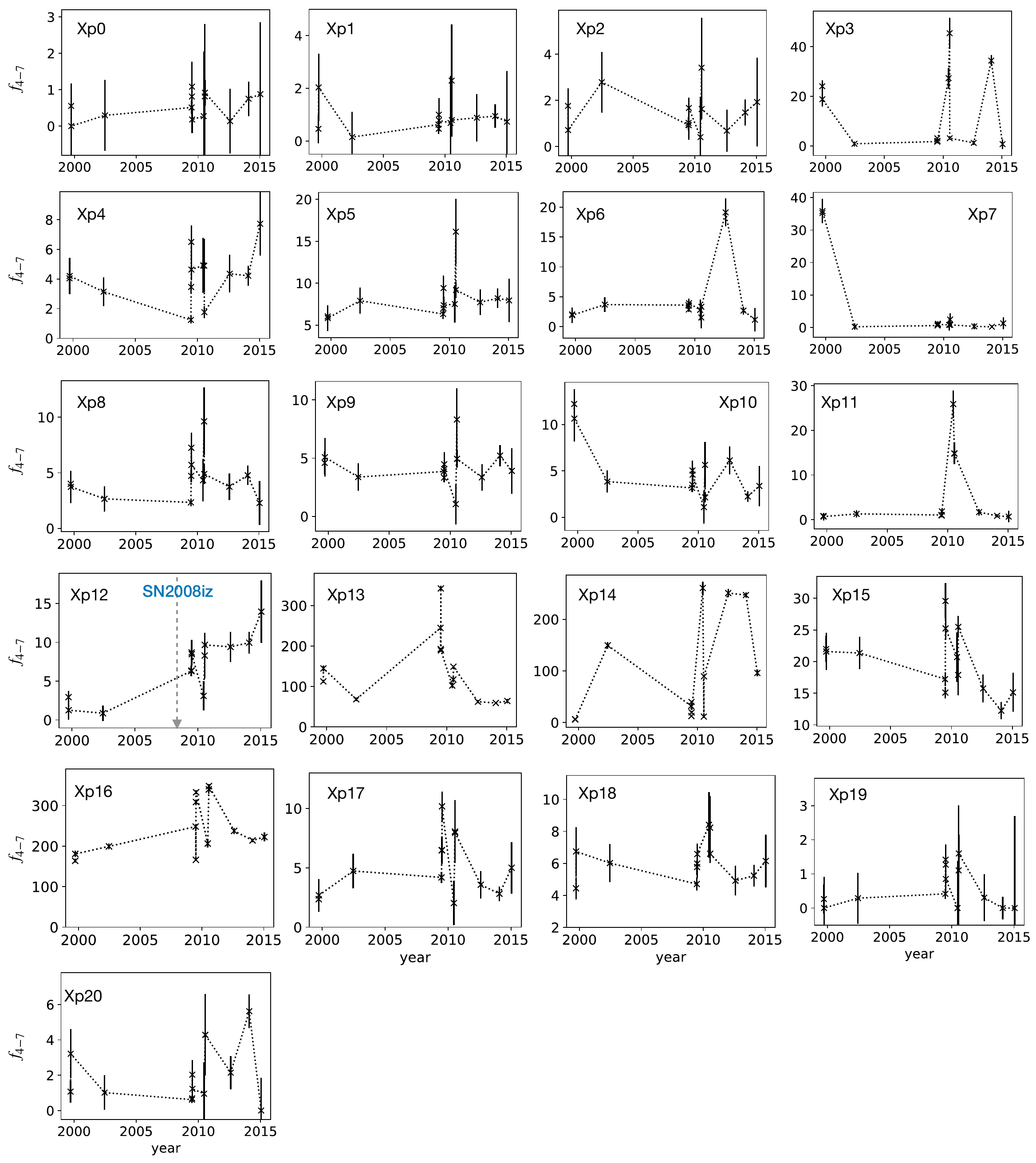}}
\caption{Light curves of 21 sources obtained from the 13 exposures in the 4-7 keV band. Fluxes are in units of $10^{-14}$ \ergpspsqcm. We note that the spectra shown in Fig \ref{fig:fn6} were taken from the ACIS-S observations from 2002 onwards.} 
\label{fig:lc}
\end{figure*}

Light curves of individual sources over the 13 observations in 1999-2015 are shown in Fig. \ref{fig:lc}. The 4-7 keV flux, $f_{4-7}$, of each source in each ObsID was obtained by integrating the low-resolution flux-density spectrum (e.g. Fig. \ref{fig:xp12spvar}) in the band. These light curves are supporting material for following purposes: 1) to discard obvious flaring sources that are clearly not SNRs (Sect. 3.2); 2) to examine long-term X-ray flux variations of selected SNR candidates (Sect. 3.5, 3.7); and 3) to argue against the apparent variability of Xp17 (Sect. 3.7).

To identify sources of transient flaring and variable sources, we used following three statistics to characterise the 13 flux measurements of each source: 1) median flux, $\tilde{f}_{4-7}$; 2) interquartile range (IQR) as a measure of variability; and 3) median of flux measurement errors, $\widetilde{\Delta f}$ (Table \ref{tab:var}).
We used median statistic and IQR, both of which are robust against outliers. In the presence of a small number of strong flaring, they represent typical values of a source in quiescence. 

Supposing that 13 individual fluxes and their errors of given light curve are $f_{i}$ and $\Delta f_{i}$ ($i=0,1,...,12$), if there is one or more $f_{i} > 5\times \tilde{f}_{4-7}$ and $(f_{i}-\tilde{f}_{4-7}) >3\times\Delta f_{i}$, or $3 \sigma $ excess, a source is considered to have transient flaring (as labelled by 'F' in Table \ref{tab:var}). With these criteria, five sources, Xp3, 6, 7, 11, and 14, are classified as transient flaring sources.

As variability increases, the IQR stretches. However, variability can be asserted only when it exceeds the fluctuations caused by random noise. If a source has a constant flux, the distribution of data is Gaussian and IQR $\approx 1.35\sigma$, where $\sigma $ is the Gaussian dispersion which we approximate by $\widetilde{\Delta f}$. Taking some margin, we consider a source as variable (labelled by 'V') when IQR $> 2\times \widetilde{\Delta {f}}$. Apart from the bright X-ray binaries (Xp13 and Xp16), four of the SNR candidates (Xp10, Xp12, Xp15, and Xp17) show signs of variability. The variabilities of Xp4, 9, and 18 are marginal as their IQR/$\widetilde{\Delta f}$ lie in the range of 1.35-2. The description of these three in the context of SNR evolution can be found in Sects. 3.5.1 and 3.7. Regarding Xp17, its light curve is highly correlated with that of bright Xp16 (M82 X-1), which is located at $\simeq 2\farcs 7$ to the north. Given the brightness and separation, as well as significant photon pile-up, it is likely that a fraction of the light from Xp16 at the PSF wing that spills over the Xp17 aperture is the cause of the apparent variability of Xp17. Xp15 is located $\simeq 4\arcsec $ from Xp16 and the spikes in the light curve, which coincide the Xp16 flaring, suggests a similar effect of Xp16 contribution. In discounting those spikes, the decreasing trend of light curve becomes clearer. The effect on the Xp15 spectrum is expected to be minor, as it is a relatively bright source.

The remaining sources with IQR$\leq 2\times \widetilde{\Delta {f}}$ and no flaring behaviour are considered to have no significant variability and labelled as 'S'. X-ray variability of some of these sources was studied by \citet{chiang11} which can be compared with these results.

\begin{table}
\caption{Flux variability}
\label{tab:var}
\centering
\begin{tabular}{rcccc}
  \hline\hline
  Xp & $\tilde{f}_{4-7}$ & IQR & $\widetilde{\Delta f}$ & class \\
  \hline
  0 & 0.55 & 0.54 & 0.69 & S  \\
  1 & 0.78 & 0.32 & 0.63 & S \\
  2 & 1.47 & 0.84 & 0.76 & S \\
  3 & 3.08 & 22.3 & 0.98 & F \\
  4 & 4.22 & 1.44 & 0.97 & S \\
  5 & 7.72 & 1.17 & 1.50 & S \\
  6 & 2.87 & 1.55 & 0.89 & F \\
  7 & 0.81 & 0.68 & 0.90 & F \\
  8 & 4.33 & 1.15 & 1.14 & S \\
  9 & 4.07 & 1.57 & 1.03 & S \\
  10 & 3.85 & 2.48 & 1.20 & V \\
  11 & 1.05 & 1.00 & 0.76 & F \\
  12 & 8.29 & 6.32 & 1.38 & V \\
  13 & 117 & 121 & 4 & V \\
  14 & 40 & 137 & 2.5 & F \\
  15 & 20.7 & 6.24 & 2.19 & V \\
  16 & 222 & 109 & 6 & V \\
  17 & 4.73 & 3.68 & 1.17 & V$^{\star}$ \\
  18 & 6.02 & 1.37 & 0.88 & S \\
  19 & 0.30 & 1.11 & 0.59 & S \\
  20 & 1.23 & 2.26 & 0.94 & S \\
  \hline
\end{tabular}
\tablefoot{Median 4-7 keV flux $\tilde{f}_{4-7}$, IQR, typical error of flux measurement approximated by their median $\widetilde{\Delta f}$. Variability class: 'S': stable or no significant variability; 'F': transient flaring; 'V': variable. $^{\star}$This is a false detection of variability due to spilling photons of a nearby, bright source Xp16.}
\end{table}

\section{Xp2: Another possible X-ray SNR}

\begin{figure}
\centerline{\includegraphics[width=0.28\textwidth,angle=0]{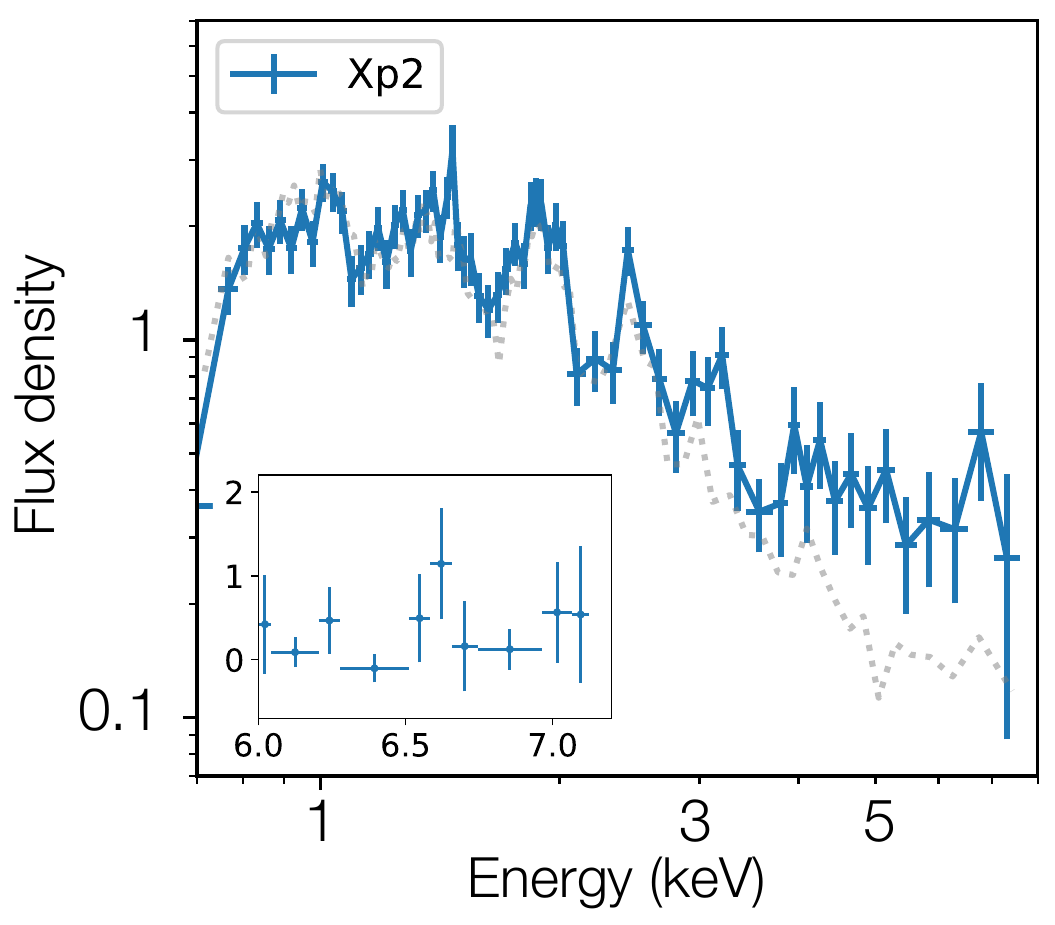}}
\caption{Flux density spectrum of Xp2, obtained from the {\it Chandra} ACIS-S. Data are in units of $10^{-14}$ erg\thinspace cm$^{-2}$\thinspace s$^{-1}$\thinspace keV$^{-1}$. Estimated diffuse background is indicated by the dotted grey line. The inset shows the Fe K band spectrum.} 
\label{fig:xp2spec}
\end{figure}

The faint X-ray source Xp2 was selected as an X-ray SNR candidates but classified as uncertain due to small detected counts in the Fe K line (Sect. 3.3). However, it is identified as a radio SNR $45.17+61.2$ and likely another X-ray SNR (Sect. 3.6, Table \ref{tab:radio}). The X-ray spectrum obtained from the Xp2 aperture is shown in Fig. \ref{fig:xp2spec}. The diffuse background dominates at lower energies and excess emission due to Xp2 emerges above 3 keV. The 4-8 keV band spectrum of Xp2 shows a Fe K line excess even after correcting for the diffuse background but with excess counts of only 5. A Gaussian fit gives the line centroid of $6.63 ^{+0.04}_{-0.05}$ keV and the line EW of $0.7\pm 0.4$ keV. The 4-8 keV flux was estimated to be $f_{4{\rm -}8} = (8.6\pm 2.1)\times 10^{-15}$ \ergps, corresponding to $L_{4{\rm -}8}=(1.3\pm 0.3)\times 10^{37}$ \ergps.

\end{appendix}

\end{document}